\documentclass[a4paper,11pt]{article}
\usepackage[small]{titlesec}
\usepackage[backend=biber,style=numeric,natbib=false,giveninits=true,doi=true,isbn=false,url=false,date=year]{biblatex}
\DeclareNameAlias{default}{last-first}

\DeclareFieldFormat[article]{title}{#1}
\renewbibmacro{in:}{}
\DeclareFieldFormat[article]{pages}{#1}
\DeclareFieldFormat{journal}{#1}
\renewcommand\bf\bfseries

\renewbibmacro*{volume+number+eid}{%
	\printfield{volume}%
	\setunit*{\addnbspace}
	\printfield{number}%
	\setunit{\addcomma\space}%
	\printfield{eid}}
\DeclareFieldFormat[article]{number}{\mkbibparens{#1}}
\DeclareFieldFormat[article]{volume}{\textbf{#1}}
\DeclareFieldFormat{year}{\mkbibparens{#1}}
\DeclareBibliographyDriver{article}{%
	\printnames{author}:%
	\newunit\newblock
	\printfield{title}%
	\newunit\newblock
	\printfield{journaltitle}
	\newunit
	\iffieldundef{number}{\printfield{volume}}{\printfield{volume}\addspace\printfield{number}}
	\addcomma\addspace\printfield{pages}\addspace
	\printfield{year}
}
\usepackage{biblatex-files}
\AtEveryBibitem{\clearfield{month}}
\AtEveryBibitem{\clearfield{day}}
\usepackage[english]{babel}
\usepackage[latin1]{inputenc}
\usepackage[T1]{fontenc}
\usepackage{csquotes}
\usepackage{bbm}
\usepackage{amsmath,amsfonts,amsthm,amsbsy,amssymb,dsfont,stmaryrd}
\usepackage[svgnames,x11names]{xcolor}
\usepackage{braket}

\usepackage{mathtools}

\numberwithin{equation}{section}

\usepackage[left=2.5cm,right=2.5cm,top=2.5cm,bottom=2.5cm]{geometry}
\usepackage[unicode=true,pdfusetitle,
bookmarks=true,bookmarksnumbered=false,bookmarksopen=false,
breaklinks=false,pdfborder={0 0 1},backref=false,colorlinks=true,linkcolor=blue,citecolor=red]
{hyperref}
\usepackage[nameinlink]{cleveref}

\usepackage{tikz}
\usetikzlibrary{arrows}
\usetikzlibrary{intersections}

\usepackage{graphicx}
\usepackage{caption}

\definecolor{ct_black}{HTML}{000000}
\definecolor{ct_orange}{HTML}{ED872D}
\definecolor{ct_purple}{HTML}{7A68A6}
\definecolor{ct_blue}{HTML}{348ABD}
\definecolor{ct_turquoise}{HTML}{188487}
\definecolor{ct_red}{HTML}{E32636}
\definecolor{ct_pink}{HTML}{CF4457}
\definecolor{ct_green}{HTML}{467821}

\definecolor{ct2_green}{HTML}{9FF781}
\definecolor{ct2_green_dark}{HTML}{088A08}

\theoremstyle{plain}
\newtheorem{thm}{\protect\theoremname}[section]
\theoremstyle{plain}
\newtheorem{lem}[thm]{\protect\lemmaname}
\theoremstyle{plain}
\newtheorem{cor}[thm]{\protect\corollaryname}
\theoremstyle{plain}
\newtheorem{prop}[thm]{\protect\propositionname}
\theoremstyle{remark}

\newtheorem{assumption}{\protect\assumptionname}[section]

\theoremstyle{remark}
\newtheorem{rem}{\protect\remarkname}[section]
\theoremstyle{definition}
\newtheorem{defn}{\protect\definitionname}[section]
\theoremstyle{plain}

  \providecommand{\assumptionname}{Assumption}
\providecommand{\claimname}{Claim}
\providecommand{\corollaryname}{Corollary}
\providecommand{\definitionname}{Definition}
\providecommand{\lemmaname}{Lemma}
\providecommand{\propositionname}{Proposition}
\providecommand{\remarkname}{Remark}
\providecommand{\theoremname}{Theorem}
\providecommand{\examplename}{Example}

\crefname{section}{Section}{Sections}
\crefname{appendix}{Appendix}{Appendices}
\crefname{figure}{Figure}{Figures}
\crefname{assumption}{Assumption}{Assumptions}
\crefname{thm}{Theorem}{Theorems}
\crefname{lem}{Lemma}{Lemmas}
\crefformat{equation}{(#2#1#3)}

\crefrangelabelformat{equation}{(#3#1#4--#5#2#6)}

\crefmultiformat{equation}{(#2#1#3}{, #2#1#3)}{#2#1#3}{#2#1#3}
\Crefmultiformat{equation}{(#2#1#3}{, #2#1#3)}{#2#1#3}{#2#1#3}

\newtheorem*{lem*}{\protect\lemmaname}

\newcommand{\norm}[1]{\left\lVert#1\right\rVert}
\DeclareMathOperator*{\slim}{s-lim}
\newcommand{\ee}{{\mathrm e}}
\newcommand{\ii}{{\mathrm i}}
\newcommand{\dd}{{\mathrm d}}
\newcommand{\dif}{\mathrm{d}}

\newcommand{\tr}{\operatorname{tr}}

\renewcommand{\Im}{\operatorname{\mathbb{I}\mathbbm{m}}}

\newcommand{\ve}{\varepsilon}

\newcommand{\im}{\operatorname{im}}

\newcommand{\SO}{S^1}
\newcommand{\BLO}{\mathcal{B}(\mathcal{H})}

\newcommand{\HB}{H}
\newcommand{\HHB}{\mathcal{H}}
\newcommand{\UB}{U}
\newcommand{\SUBO}{\sigma(\UB(\T))}

\newcommand{\HHE}{\mathcal{H}_\mathrm{E}}
\newcommand{\HE}{H_\mathrm{E}}
\newcommand{\UE}{U_\mathrm{E}}

\newcommand{\T}{1}

\newcommand{\Id}{\mathds{1}}

\newcommand{\IB}{\mathcal{I}}
\newcommand{\IE}{\mathcal{I}_\mathrm{E}}

\newcommand{\MB}{M}
\newcommand{\MMB}{\mathcal M}
\newcommand{\MME}{\mathcal P_\mathrm{E}}

\newcommand{\VB}{V}
\newcommand{\VE}{V_\mathrm{E}}

\newcommand{\eab}{\ve_{\alpha\beta}}

\newcommand{\EVS}{\mathcal E} 

\title{Strongly Disordered Floquet Topological Systems}
\author{Jacob Shapiro and Cl\'ement Tauber \\
\footnotesize{Institute for Theoretical Physics, ETH Z\"{u}rich}} 
\date{\vspace{-0.5cm}}

\begin{document}
	
\maketitle

\begin{abstract}
We study the strong disorder regime of Floquet topological systems in dimension two, that describe independent electrons on a lattice subject to a periodic driving. In the spectrum of the Floquet propagator we assume the existence of an interval in which all states are localized--a mobility gap. First we generalize the relative construction from spectral to mobility gap, define a bulk index for an infinite sample and an edge index for the half-infinite one and prove the bulk-edge correspondence. Second, we consider completely localized systems where the mobility gap is the whole circle, and define alternative bulk and edge indices that circumvent the relative construction and match with quantized magnetization and pumping observables from the physics literature. Finally, we show that any system with a mobility gap can be reduced to a completely localized one. All the indices defined throughout are equal.
\end{abstract}

\section{Introduction}

In the context of topological insulators, a Floquet system describes independent electrons on a lattice subject to a periodic driving beyond the adiabatic regime. The system is topological when one can define a stable index that captures some topological property of the sample, either in the bulk of an infinite one or at the edge of a semi-infinite one. For Floquet systems, the latter is sometimes associated to a transport observable, and usually coincides with the bulk index (whose physical meaning is sometimes associated to magnetization, see below) through the celebrated bulk-edge correspondence \cite{RudnerPRX13,SadelSchulz-Baldes17,GrafTauber18}. Originally designed to induce topological properties on a trivial sample through the periodic driving \cite{OkaAoki09}, Floquet topological systems have recently become a topic of intense study when it was realized that this driving also allowed to engineer new topological phases of matter that have no static counterpart \cite{RudnerPRX13}; some proposals for experimental observation of these phases in cold atoms were recently suggested \cite{NathanPRL16,Quelle_etal17}. 

So far the main prerequisite to define topological indices in Floquet systems has been the presence of a gap in the spectrum of the unitary Floquet propagator, describing time evolution in the bulk after one period of driving. In this context the bulk-edge correspondence was first established in clean systems and then extended to weakly disordered samples, for various dimensions and symmetries \cite{RudnerPRX13,Lyon15,Fruchart16,FulgaMaksymenko16,GrafTauber18,SadelSchulz-Baldes17,LiuHarperRoy18}. By analogy with static systems, the effect of disorder is to progressively fill the spectral gap of the propagator by localized states \cite{TitumPRX16,HamzaJoyeStolz09}, and all the previous results work only as long as the spectral gap remains open. 

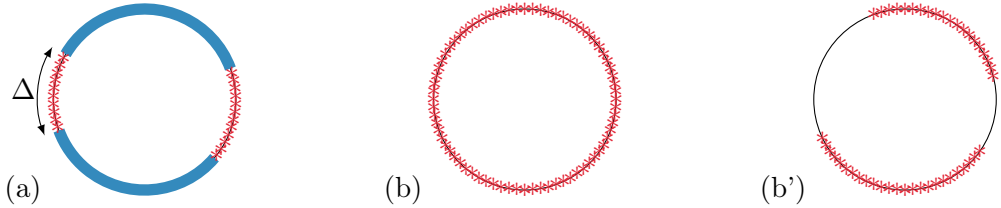
\begin{figure}[htb]
	\centering
	\begin{tikzpicture}
	\newdimen\r
	\pgfmathsetlength\r{1.2cm}
	\draw[] (0,0) circle (\r);
	\draw[line width=0.15cm,ct_blue] (20:\r) arc (20:150:\r);
	\draw[line width=0.15cm,ct_blue] (-40:\r) arc 
	(-40:-160:\r);
	\foreach \k in {1,...,10}
	{\draw[ct_red] (148 + 5*\k:\r) node{${\ast}$};
	}
	\foreach \k in {1,...,12}{
		\draw[ct_red] (22 - 5*\k:\r) node{$\ast$};
	}
	\draw[latex-latex] (150:1.4) arc (150:200:1.4);
	\draw (-1.6,-\r) node {(a)};
	\draw (175:1.6) node {$\Delta$};
	
	\begin{scope}[xshift=5cm]
	\newdimen\r
	\pgfmathsetlength\r{1.2cm}
	\draw[] (0,0) circle (\r);
	\foreach \k in {1,...,72}
	{\draw[ct_red,thick] (5*\k:\r) node{${\ast}$};
	}
	\draw (-1.6,-\r) node {(b)};
	\end{scope}
	
	\begin{scope}[xshift=10cm]
	\newdimen\r
	\pgfmathsetlength\r{1.2cm}
	\draw[] (0,0) circle (\r);
	\foreach \k in {1,...,20}
	{\draw[ct_red,thick] (10+5*\k:\r) node{${\ast}$};
	}
	\foreach \k in {1,...,25}
	{\draw[ct_red,thick] (-30-5*\k:\r) node{${\ast}$};
	}
	\draw (-1.6,-\r) node {(b')};
	\end{scope}
	\end{tikzpicture}
	\caption{Examples of spectrum for the Floquet propagator $\UB(\T)$. (a) Typical situation with a mobility gap $\Delta$ (possibly several), represented by red crosses. The remaining part of the spectrum, in blue, can be arbitrary. For completely localized operators, the mobility gap is the entire circle (b) or possibly with spectral gaps (special case of a mobility gap) (b'). The relative construction applies to all cases, but can be avoided in (b) and (b') through magnetization and pumping indices. The stretch-function construction maps situation (a) to (b) by stretching $\Delta$ onto $\SO\setminus\{1\}$ and mapping the remainder of the spectrum to $1$. \label{fig:summary}}
\end{figure}

This paper deals with two-dimensional systems with no particular symmetry (class A of \cite{AltlandZirnbauer97}). We address the problem of strong disorder, when the gap is completely filled by localized states (see \cref{fig:summary}). This is the so-called mobility gap regime, that is characterized by Anderson localization, and mathematically through the fractional moment condition \cite{Aizenman_Graf_1998}. The starting point of our work is a general and almost-sure consequence of this condition, which we take as a deterministic assumption to define the mobility gap. This approach is analogous to the other few works on topological properties of strongly disordered systems in the static case, first studied for the Integer Quantum Hall Effect \cite{EGS_2005} and more recently in chiral systems \cite{Graf_Shapiro_2018_1D_Chiral_BEC}. Moreover the fractional moment condition has been already established for unitary random operators \cite{HamzaJoyeStolz09,AschBourgetJoye10}, as well as some numerical evidence of localization in Floquet topological models \cite{TitumPRX16}. We note in passing that \cite{PRODAN20161150} also studied strongly-disordered unitary topological systems in the bulk, however, they used a covariant probabilistic framework.

The first result of this paper is to show that the so-called relative construction, developed in the spectral gap case in dimension two \cite{RudnerPRX13,SadelSchulz-Baldes17,GrafTauber18}, can be extended to the mobility gap regime. This construction reduces the physical unitary evolution to a time periodic propagator in the bulk, which has a well-defined index. This requires a logarithm of the Floquet propagator, that we prove to be well-defined with a branch-cut in the mobility gap. Thanks to the estimates coming from localization, the logarithm is weakly-local--its matrix elements in the position basis have rapid off-diagonal decay, and possible diagonal blowup. With this we can adapt the proof in \cite{GrafTauber18} of the bulk-edge correspondence from the spectral gap case, in which the Combes-Thomas estimate was used instead of localization.

The physical implementation of this relative construction is however not straight-forward, and one can look for situations where it may be circumvented. For clean samples, a Floquet system is actually an insulator only when the Floquet propagator is exactly the identity operator $\Id$, for which the relative construction is not required. In the spectral and mobility gap cases the system is not insulating anymore and the relative constructions somehow subtracts the other transport contributions from the topological one \cite{Tauber18}. In the strongly disordered case the analogue of $\Id$ is to consider a Floquet propagator that is completely localized, namely that its entire spectrum is a mobility gap (see \cref{fig:summary}(b) and (b')). It was shown in \cite{TitumPRX16} that in contrast to the static case \cite{EGS_2005}, such systems may still have edge modes and topological properties. Moreover the indices can be computed without the relative construction and have a nice physical interpretation in terms of quantized orbital magnetization in the bulk \cite{NathanPRL16} and quantized pumping at the edge \cite{TitumPRX16}. The second result of this paper is a rigorous definition of these indices and a proof of their respective bulk-edge correspondence.

Our last result is to show that any mobility gap situation can actually be reduced to a fully localized case, for which the previous indices can be used and circumventing again the relative construction. This reduction is done through the smooth functional calculus with a particular function that stretches the mobility gap onto the entire circle (as in \cite{SadelSchulz-Baldes17} who used this construction only for the edge in the spectral gap regime). 

We finally show that the indices defined in this approach coincide with the ones of the relative construction (which are also defined in this case). To that end we show the continuity of the bulk relative index along a specific path of deformation. We believe this continuity result in the mobility gap regime is important because it joins an extremely short list of results: the deterministic constancy of the quantum Hall conductivity w.r.t. the Fermi energy proven in \cite{EGS_2005}. Thus \cref{thm:F_invariance} opens interesting perspectives in the investigation of the topology of deterministic mobility gapped systems, of which very little is known.

Finally note that quantum walks, namely finite sequences of unitary operators, can also be seen as discrete-time Floquet systems, for which topological indices have been already defined in clean and weakly disordered models \cite{TauberDelplace15,AschBourgetJoye17}. In some cases the Floquet formalism can be applied to quantum walks \cite{DelplaceFruchartTauber17,SadelSchulz-Baldes17}, so that our result should in principle cover the strongly disordered version of these quantum walks.

The paper is organized as follows. After describing the setting and stating the three results mentioned above in \cref{sec:setting}, we detail their respective proofs in \cref{sec:RelativeBEC,sec:completely_localized,sec:stretch_function}. Finally, \cref{app:RAGE,app:HJ_formula,app:convergence_lemma} are results of independent interest for Floquet systems or more general unitary operators.

\medskip
\noindent\textbf{Acknowledgements:} The authors thank Gian Michele Graf for many useful discussions. J.S. thanks Netanel H. Lindner for his hospitality at the Technion in Israel and for elaborations on the completely localized edge index. C.T thanks Alain Joye for his invitation at institut Fourier and fruitful discussions on localization for unitary operators.

\section{Setting and main results \label{sec:setting}}

Let a time-dependent periodic Hamiltonian $\HB:\SO\to\BLO$ be given where $\HHB:=\ell^2(\mathbb{Z}^d)\otimes\mathbb{C}^N$ is the (bulk) Hilbert space and $N,d\in\mathbb{N}_{\geq1}$ are the (fixed) number of degrees of freedom per lattice site and the space dimension. Here we use $\SO\cong[0,1]/\{0\sim1\}$. We assume the following conditions about $H$ throughout:

\begin{assumption}\label{assu:continuity of Hamiltonian} (\emph{Continuity})
	$t\mapsto H(t)$ is strongly-continuous except for a finite number of jump discontinuities.
\end{assumption}
In what follows, let $(\delta_x)_{x\in\mathbb{Z^d}}$ be the canonical (position) basis of $\ell^2(\mathbb{Z}^d)$ and the map $H(t)_{xy}=\langle\delta_x,H(t)\delta_y\rangle:\mathbb{C}^{N}\to\mathbb{C}^{N}$ acts between the internal spaces of $x$ and $y$; $\|\cdot\|$ is the trace norm of such maps and also the $1$-norm on $\mathbb{R}^d$ when applied to some $x\in\mathbb{Z}^d$: $\|x\|\equiv\sum_{i=1}^{d}|x_i|$.
\begin{assumption}\label{assu:locality of Hamiltonian} (\emph{Locality}) There are some constants $C,\mu>0$ such that for any $t\in\SO$
we have 
	\begin{align}
	 \|H(t)_{xy}\| < C \ee^{-\mu\|x-y\|}\qquad(x,y\in\mathbb{Z}^d)\,.\label{eq:locality of Hamiltonian}
	\end{align}	
\end{assumption}
We use the symbols $C,\mu$ for constants which may change from line to line, but which are otherwise independent of the other variables unless stated differently.

To deal with Floquet systems we consider the unitary propagator $\UB:[0,1]\to\BLO$ generated by $\HB$, that is, the unique solution to $\ii \dot{\UB} = \HB \UB$ with initial condition $\UB(0) = \Id$ (it is a fact that even though $\HB$ is periodic, $\UB$ need not be and so its domain is a-priori $[0,1]$). A well-defined topological phase exists depending on spectral or dynamical properties that $\UB(\T)$ may or may not satisfy. Such a phase was established (see e.g. \cite{RudnerPRX13,SadelSchulz-Baldes17,GrafTauber18}) in the presence of a spectral gap:

\begin{defn} (\emph{Spectral gap}) $\UB(\T)$ has a spectral gap iff its spectrum is not the entire circle: \begin{align} \SUBO \neq \SO\,. \label{eq:spectral gap}\end{align} Since $\SUBO$ is a closed subset of $\SO$, the existence of a point outside it implies the existence of a whole open interval outside of it, which is called \emph{a spectral gap}. In contrast to the parametrization of the domain of $H$, here we rather use $\SO\cong\{z\in\mathbb{C}\,|\,|z|=1\}$. 
\end{defn}

The main point of the present paper is that even when \cref{eq:spectral gap} doesn't hold, a weaker gap, called \emph{a mobility gap} may exist, which still allows for the definition of topological indices. The following definition of a mobility gap, by analogy with \cite{EGS_2005,Graf_Shapiro_2018_1D_Chiral_BEC} for static systems, corresponds to an almost-sure consequence of a probabilistic analysis in which (fractional moments) localization is established. Such proofs of localization for unitary models have so far been established in e.g. \cite{HamzaJoyeStolz09,AschBourgetJoye10}.

\begin{defn}\label{def:mobility gap} (\emph{Mobility gap}) The interval $\Delta\subseteq\SO$ is a mobility gap for $\UB(\T)$ iff (1) there is some constant $\mu>0$ such that for any $\ve>0$ there is some constant $0\leq C_\ve<\infty$ such that we have
	\begin{align}\label{eq:moblity_gap}\sup_{g\in B_1(\Delta)}
 \|g(\UB(1))_{x,y})\| \leq C_\ve \ee^{-\mu\|x-y\|+\ve\|x\|}\qquad(x,y\in\mathbb{Z}^d)
	\end{align}
	 where $B_1(\Delta)$ is the space of all Borel maps $g:\SO\to\mathbb{C}$ which are constant outside of $\Delta$ and obey $|g(z)|\leq1$ for all $z\in\SO$; and (2) all 
	 eigenvalues of $\UB(\T)$ within $\Delta$ are of finite degeneracy.
\end{defn}

\begin{rem}
	The supremum over $g$ implies dynamical localization within $\Delta$ by considering the family of functions $\SO\ni\lambda\mapsto \lambda^n \chi_{\Delta}(\lambda)$ indexed by $n$. Consequently $U(1)$ has pure point spectrum within $\Delta$, due to a RAGE theorem analogue for unitaries \cite{EnssVeselic83, HamzaJoyeStolz09}. This is detailed in \cref{app:RAGE}.
\end{rem}

\begin{rem}
	Ignoring the supremum over $g$, the decay condition \cref{eq:moblity_gap} is weaker than the locality condition \cref{eq:locality of Hamiltonian} due to the presence of $\ve$ which means that while the matrix elements decay in the off-diagonal direction, the rate is non-uniform in the diagonal direction. 
	We call this property \emph{weakly-local}. This is the best almost-sure consequence that probability theory can provide, as was formulated first in \cite{EGS_2005}[Eq.~(1.2)].
\end{rem}
 
\begin{rem}
	\cref{eq:moblity_gap} is actually weaker than what comes out of the fractional moments condition. One could replace the exponential weight $\ee^{\ve \|x\|}$ by the inverse $|a(x)|^{-1}$ of \emph{any} $a\in\ell^1(\mathbb{Z}^d)$; the constant $C$ then depends on $a$.
\end{rem}

\begin{rem}
	We need to require finite degeneracy for the eigenvalues in $\Delta$ in addition because here we start with a deterministic assumption, but within a probabilistic model such a zero-one law would come out as is usual for Anderson localization proofs (see \cite{Simon94}). 
\end{rem}

\begin{rem}\label{rem:spectral gap implies mobility gap}
	\cref{eq:spectral gap} implies \cref{def:mobility gap} via the Combes-Thomas estimate, so that a spectral gap is also a mobility gap and we may treat both by treating only the mobility gap.
\end{rem}

\paragraph{The edge sample.}

In the edge picture the Hilbert space is $\HHE=\ell^{2}(\mathbb N \times \mathbb Z^{d-1}) \otimes \mathbb C^N$ describing independent electrons on a half-space. The canonical embedding $\iota: \HHE \hookrightarrow \HHB$ and truncation $\iota^\ast: \HHB \twoheadrightarrow \HHE$ satisfy $\iota^\ast\iota = \Id$ on $\HHE$ and $\iota \iota^\ast = P_1$ where $P_1:=\Theta(X_1)$ is the projection in $\HHB$ onto states supported in the half-space, with $X_i$ the $i$th component of the position operator $X$ and $\Theta$ the step function. For $A$ acting on $\HHB$ we denote the corresponding truncated operator on $\HHE$ by $\widehat A := \iota^\ast A \iota$. In particular the edge Hamiltonian is
\begin{align}
\HE(t) := \widehat{\HB(t)}
\end{align}
corresponding to Dirichlet boundary condition, although other conditions could be implemented in principle. $\HE$ inherits some properties of $\HB$, in particular it satisfies \cref{eq:locality of Hamiltonian}, and generates a unitary propagator $\UE$ on $\HHE$ through $\ii \dot{\UE} = \HE \UE$ and $\UE(0)=\Id$. All these properties rely only on the fact that $\HB$ is local by \cref{eq:locality of Hamiltonian} and not on the existence of (any) gap of $\UB(\T)$, so they remain true in the mobility gap regime.

In what follows the topological indices are defined through the use of switch functions $\Lambda : \mathbb Z \rightarrow \mathbb R$ such that $\Lambda(n) = 1$ (resp. $0$) for $n$ large and positive (resp. negative).
We denote by $\Lambda$ the corresponding  multiplication operator on $\ell^{2}(\mathbb Z)$, and by $\Lambda_i=\Lambda(X_i)$ a switch function in direction $i$ acting on $\ell^2(\mathbb Z^d)$ or $\ell^2(\mathbb N \times \mathbb Z^{d-1})$. 

\subsection{The relative construction}

Here we finally specify to the case $d=2$ and no symmetry. For this case, the bulk-edge correspondence is established in \cite{GrafTauber18} first when $\UB(\T)=\Id$ (so $\SO\setminus\{1\}$ is a "special" spectral gap) and then when $\UB(\T)$ has a general spectral gap. The latter case was reduced to the first one by constructing a relative evolution, generated by an effective Hamiltonian. It turns out that the same procedure can be followed in the mobility gap regime. The effective Hamiltonian is defined through a logarithm of the one-period propagator
\begin{align}\label{eq:effective Hamiltonian}
H_\lambda := \ii \log_\lambda(\UB(\T))
\end{align}
where $\lambda \in \Delta$ is chosen inside the mobility gap and used as a branch cut for the (principal) logarithm. We show in \cref{cor:effective Hamiltonian and its evolution are weakly-local} that $H_\lambda$ is \emph{weakly-local} as a weaker property of being local (i.e. satisfying \cref{eq:locality of Hamiltonian}). Note that in the spectral gap case the discontinuity of the logarithm, which is otherwise analytic, maybe ignored since it occurs out of the spectrum of $\UB(\T)$; since analytic functions of local operators are local \cite[Appendix D]{Aizenman_Graf_1998}, the logarithm is local too. For us, however, $H_\lambda$ is merely weakly-local via localization. This is enough to define the indices, as we shall see.

For two operators $A,B:[0,\T]\to\BLO$, not necessarily periodic, we define the concatenation in time $A\#B:[0,\T]\to\BLO$ by
\begin{align}\label{time_concatenation}
(A\#B)(t) = \left\lbrace \begin{array}{ll}
A(2t), & (0 \leq t \leq \T/2) \\
B(2(\T-t)), &  (\T/2 \leq t \leq  \T)
\end{array}\right..
\end{align}
The two operators occur consecutively in time, the second backwards. The relative bulk Hamiltonian is then defined by $\HB^\mathrm{rel} := 2(\HB\#H_\lambda)$. The effective Hamiltonian being time-independent, its unitary propagator is $U_\lambda(t) = \ee^{-\ii t H_\lambda}$ and satisfies $U_\lambda(\T) = \UB(\T)$ by construction. It follows that the relative evolution generated by $\HB^\mathrm{rel}$, $\UB^\mathrm{rel} = \UB \# U_\lambda$, satisfies $\UB^\mathrm{rel}(1) = \Id$. Similarly, the relative edge Hamiltonian is defined by $\HE^\mathrm{rel} := 2(\widehat \HB\# \widehat {H_\lambda}) = \widehat {\HB^\mathrm{rel}}$, which generates $\UE^{\mathrm{rel}}$. Note that $\widehat{H_\lambda}$ is the truncation of $H_\lambda$ and \emph{not} the logarithm of $\UE(\T)$ for which we don't assume a mobility gap to exists. In particular $\UE(\T) \neq \ee^{-\ii \widehat {H_\lambda}}$ so that $\UE^{\mathrm{rel}}$ is not given by their concatenation.

Finally recall the non-commutative three-dimensional winding $W$ defined for a unitary loop $V:\SO\to\BLO$ is given by the formula \begin{align}
 W(V) \equiv -\frac{1}{2}\int_0^1 \tr  \dot{V} V^\ast[V_{,1}V^\ast, V_{,2} V^\ast]\,.\label{eq:Cartan-Mauerer-form}
\end{align}
Here and in what follows, we use the short-hand notation for the non-commutative spatial derivative in direction $j$ of an operator $A$ as $A_{,j}:=\partial_j A \equiv \ii[\Lambda_j,A]\equiv\ii[\Lambda(X_j),A]$. \cref{eq:Cartan-Mauerer-form} is equivalent to a pairing between the  $K_1$ class defined by $V$ and a three-dimensional Chern character, as detailed \cite{PRODAN20161150}.

We come to the main result of this paper

\begin{thm}\label{thm:BEC_MG}
 Under \cref{assu:continuity of Hamiltonian,assu:locality of Hamiltonian} and additionally assuming \cref{def:mobility gap} holds for $\UB(1)$: (1) the bulk index $\IB \equiv W(\UB^\mathrm{rel})$ is finite, integer-valued, and independent of the choices of switch functions and branch cut $\lambda\in\Delta$; (2) the edge index
 \begin{align}\label{eq:IE_MG}
 \IE \equiv -\ii \tr \UE^\mathrm{rel}(\T)^\ast  \partial_2\UE^\mathrm{rel}(\T)
 \end{align}
 is finite, integer-valued and independent of the choice of switch function; (3) the bulk-edge correspondence holds
 \begin{align}
\IB = \IE\,.
 \end{align}
\end{thm}

If $H_\lambda$ were local, then so would be $U_\lambda$, $\UB^\mathrm{rel}$ and $\UE^\mathrm{rel}$ and this theorem would be already covered by \cite[Theorem 3.8]{GrafTauber18}. Here instead we need to adapt the proof to weakly-local operators. In particular we need to show that $\partial_j A$ also have a so-called confining property when $A$ is weakly-local, so that expressions involved in \cref{eq:Cartan-Mauerer-form} and \cref{eq:IE_MG} are trace-class. Apart from that point, the other properties of $\IB$ and $\IE$ as well as the bulk-edge correspondence follow the same route as in \cite{GrafTauber18}. The proof of \cref{thm:BEC_MG} is detailed in \cref{subsec:The relative BEC}.

In conclusion, in contrast to the integer quantum Hall effect (IQHE henceforth) where the mobility gap bulk-edge correspondence is quite different in the spectral and mobility gap regime (cf. \cite{Elbau_Graf_2002} vs. \cite{EGS_2005}), the relative construction works similarly for both cases in Floquet topological insulators, once one generalizes from local to weakly-local operators and shows the desired properties of the discontinuous logarithm, \cref{cor:effective Hamiltonian and its evolution are weakly-local}. 

\begin{rem}
	A possible objection to the relative construction is the following. In defining $\IE$, the truncated generator of the bulk relative propagator, $\HE^\mathrm{rel}$ (which depends on the logarithm of the \emph{bulk} evolution), and not just the truncated bulk Hamiltonian, $H_E$, has been used, so that \cref{thm:BEC_MG} actually connects between $\IB$ and an edge index which contains significant information from the bulk. It is thus legitimate to ask for a more independent formulation where bulk and edge indices are strictly separated already at the level of the Hamiltonians, without intertwining their evolutions. The conclusion is however that the relative construction is perfectly valid since the stretch-function construction coincides with it in the end, as we show in \cref{thm:F_invariance}.
\end{rem}

\begin{rem}
	Even if the system has a gap, it is of interest to probe the system when placing the branch cut of the logarithm within the localized spectrum, in analogy with the explanation of the plateaus of the IQHE. Thus, if one day we could experimentally determine the position of the branch cut, our results would explain the corresponding plateaus which will be measured in $\IB$.
\end{rem}
\subsection{Completely localized systems} Even though the relative construction works both in the spectral and mobility gap regime, it is still legitimate to wonder if it is actually necessary. Interestingly, the formula \cref{eq:Cartan-Mauerer-form} is finite also for unitary maps which are \emph{not} periodic (whose domain is $[0,1]$ rather than $\SO$), but is an integer (and hence possibly an index) only when the map is periodic \cite[Proposition 3.3]{GrafTauber18}, \emph{e.g.} for the relative evolution $\UB^\mathrm{rel}$ which by construction has $\UB^\mathrm{rel}(\T)=\UB^\mathrm{rel}(0)=\Id$. In this section we propose an alternative definition of the bulk index when the physical evolution is not periodic, $\UB(\T)\neq\Id$, in order to avoid the relative construction. Instead we assume that $\UB(\T)$ is \emph{completely localized}. In order to define this notion, let us recall that given a self-adjoint projection $P$ on $\HHB$, its Chern number \cite{EGS_2005} is defined by \begin{align}c(P) \equiv 2 \pi \ii \tr P[\partial_1 P,\partial_2 P]  \end{align} which is finite for $P$ weakly-local in the sense of \cref{eq:moblity_gap} and zero for $P$ of finite rank. Colloquially we refer to the Chern number of an eigenvalue as the Chern number of the spectral projection onto that eigenvalue.
\begin{defn}\label{def:completely localized U} (\emph{Completely localized unitaries})
	We say that $\UB(\T)$ is \emph{completely localized} iff (1) its spectrum is one mobility gap, that is $\Delta=\SO$, in the sense of \cref{def:mobility gap} except for finitely many infinitely degenerate eigenvalues; (2) The Chern number of each of its (infinitely degenerate) eigenvalues is zero.
\end{defn}
In particular $\UB(\T)$ has only pure point spectrum. We denote by $\EVS \subset \SO$ its set of eigenvalues (a countable set) and by $P_\lambda$ the associated spectral projection onto an eigenvalue $\lambda\in\EVS$. By \cref{def:mobility gap} all but finitely many $\lambda$'s have $\dim \im P_\lambda < \infty$. Let $\alpha_1,\dots,\alpha_M$ be the eigenvalues for which this is false. \cref{def:mobility gap} ensures that  $c(P_{\alpha_i})$ makes sense, and \cref{def:completely localized U} further stipulates they should vanish.

This is the so-called anomalous phase \cite{RudnerPRX13}. Note that infinite rank projections appear in the construction of \cref{subsec:stretchfunction} below, but also in some Floquet models such as the Anomalous Floquet Anderson Insulator \cite{TitumPRX16}. The assumption on the vanishing of the Chern numbers is satisfied there, but in principle it is possible to engineer other completely localized models where it is not.

Inspired by \cite{NathanPRL16}, we define the orbital magnetization corresponding to $\UB$:
\begin{defn} (\emph{Magnetization}) For the evolution $U:[0,1]\to\BLO$ (which need not be periodic), define the magnetization operator \begin{align}\label{eq:def_MB} \MB(U) := \ \int_0^\T \Im( \UB^\ast \Lambda_1 \HB \Lambda_2 \UB) \,, \end{align}where $\Im A\equiv\frac{1}{2i}(A-A^\ast)$ is the imaginary part of $A$, and the total magnetization (a number) as 	\begin{align}
	\MMB(\UB) := \sum_{\lambda \in \EVS} \tr P_\lambda \MB(\UB) P_\lambda\,.
	\end{align} where $P_\lambda$ are the spectral projections onto the eigenvalues of $\UB(\T)$ as above.
\end{defn}
Note that the integrand in $M(U)$ can be rewritten 
\begin{align}
\Im(U^\ast \Lambda_1 H \Lambda_2 U)&=\dfrac{1}{2} \big(U^\ast \Lambda_1 U \partial_t (U^\ast \Lambda_2 U) - 1 \leftrightarrow 2 \big).
\end{align}
Pretending $\Lambda_i \sim r_i$, the position operator, the latter expression is $1/2\, \mathbf r(t) \times \dot{\mathbf r}(t)$, which up to a prefactor corresponds to the orbital magnetization. The physical aspects of $\MB$ and $\MMB$, including a proposal for an experimental realization in cold atoms, were studied in detail in \cite{NathanPRL16}.

\begin{thm}\label{thm:magnetization}
	If $\UB(\T)$ is completely localized in the sense of \cref{def:completely localized U}, the magnetization $\MMB(\UB)$
	is finite, integer-valued and independent of the choice of switch functions. Moreover $\MMB(\UB) = \IB$. If $\UB(\T) = \Id$ then $\MMB(\UB) = W(\UB)$ and if $\HB$ is time-independent then $\MMB(\UB) = 0$.
\end{thm}

Thus for completely localized systems $\UB$ the computation of the index $\MMB(\UB)$ does not require the relative construction, but the price to pay is that operator $\MB(\UB)$ is not trace-class anymore. However it is summable in the eigenbasis of $\UB(\T)$, with sum $\MMB(\UB)$. We emphasize that a mobility gap also applies when $\SUBO \neq \SO$; i.e. for a spectrally gapped system obeying \cref{def:completely localized U} the relative construction can also be circumvented using magnetization.

The strategy of the proof is to use the relative construction by choosing an effective Hamiltonian $H_\lambda$ for an arbitrary $\lambda \in \Delta$. As detailed in \cref{sec:completely_localized} we get
\begin{align}\label{eq:I=M-Mlambda}
\IB = \MMB(U) - \MMB(U_\lambda)\,.
\end{align}
The effective Hamiltonian being time-independent the magnetizations simplifies to
\begin{align}
\MMB(U_\lambda) = \sum_{z \in \SO}  \tr\big(P_z \, \ii  \big( \Lambda_1 \HB_\lambda \Lambda_2 - \Lambda_2 \HB_\lambda \Lambda_1 \big) P_z \big)
\end{align}
and the difficulty is to show that this expression is well-defined and vanishes. Note that a similar expression already appeared in the context of the IQHE as an extra term required to establish the bulk-edge correspondence of Hall conductivity in the mobility gap regime \cite{EGS_2005}. An interpretation in terms of magnetization was also proposed there for time-independent Hamiltonians. However in that case the magnetization is not vanishing because the mobility gap is not the entire spectrum.

For completely localized systems, it is also possible to define an edge index without the relative construction, also related to the previous one through the bulk-edge correspondence.

\begin{thm}\label{thm:edge_magnetization}
If $\UB(\T)$ is completely localized as in \cref{def:completely localized U}, the time-averaged charge pumping
\begin{align}
\MME(\UE(\T)) = \lim_{n\rightarrow \infty} \lim_{r\rightarrow \infty} \dfrac{1}{n}\tr \Big( ((\UE(\T))^\ast)^n[\Lambda_2, \UE(\T)^n] Q_{1,r} \Big)
\end{align}
where $Q_{1,r} = \chi_{\leq r}(X_1)$, is finite, integer valued, and independent of the choice of switch function. Moreover bulk-edge correspondence holds 
\begin{align}
\MME(\UE(\T)) &= \MMB(\UB)
\end{align}
\end{thm}

The physical interpretation of $\MME$ is a quantized pumping of charges, counted through $((\UE(\T))^\ast)^n \Lambda_2 \UE(\T)^n - \Lambda_2$, that is confined at the edge \cite{GrafTauber18,TitumPRX16}: if the corresponding $\UB(\T)=\Id$, the pumping is quantized within a single cycle, whereas for completely localized $\UB(1)$, the quantization is true on average over time only, and coincides with magnetization.

\subsection{The stretch-function construction \label{subsec:stretchfunction}}

The previous section extends the definition of bulk and edge indices beyond  $\UB(\T)=\Id$ without using the relative construction. However it only works for completely localized systems. Here we propose a way to reduce the general situation ($\Delta\neq\SO$) to the one described by \cref{def:completely localized U}.

\begin{defn} (\emph{Stretch function})
	Let $\Omega$ be an interval in  $\SO\subseteq\mathbb{C}$. A stretch function $F_\Omega:S^1 \to S^1$ is a smooth function such that $F_\Omega(z) = 1$ for $z \in S^1 \setminus \Omega$ and 
	\begin{align}
	\dfrac{1}{2\pi \ii }\int_{S^1} F_\Omega(z)^{-1} \dd F_\Omega(z) = 1\,.
	\end{align}
\end{defn}

The role of $F$ is to stretch the interval $\Omega$ onto the entire circle except the point at one, which is the image of $S^1 \setminus \Omega$. In particular if $\Omega = S^1 \setminus\{1\}$ then the identity $F_\Omega(z) = z$ is an appropriate winding function. We think of $F_\Omega$ as a function which selects the appropriate (mobility or spectral) gap, analogous to (a smooth deformation of) the function $\chi_{(-\infty,E_F)}$ with $E_F\in\mathbb{R}$ the Fermi energy, for the IQHE.  For a given interval $\Omega$ we define
\begin{align}
\VB(t) := F_{\Omega}(\UB(t)), \qquad \VE(t) := F_{\Omega}(\UE(t))\qquad(t\in[0,1])
\end{align}
via the functional calculus. $\VB$ and $\VE$ are two unitary families on $\HHB$ and $\HHE$ respectively, that satisfy $\VB(0) =\VE(0)=\Id$. Moreover if $\Omega=\Delta$ is a mobility gap of $\UB(\T)$, then by construction the entire circle $\SO$ is a mobility gap of $\VB(\T)$, so that the latter is completely localized in the sense of \cref{def:completely localized U}. Indeed, $\SO\setminus\{1\}$ is a mobility gap for $\VB(\T)$ in the sense of \cref{def:mobility gap} and $\{1\}$ may be an infinitely degenerated eigenvalue which however necessarily has vanishing Chern number by additivity, since both Chern numbers of ${\chi_{\SO}(\VB(\T))}=\Id$ and ${\chi_{\SO\setminus\{1\}}(\VB(\T)) }$ vanish, as the latter contains only finite rank projections.

\begin{cor}\label{cor:BEC_stretched_operators}
	If $\Delta$ is a mobility gap of $\UB(\T)$ and $F_\Delta$ is a stretch function, $\VB(\T)$ is completely localized so that  
	\begin{align}
	\IB':= \MMB(\VB),\qquad \IE':= \MME(\VE(1))
	\end{align}
	are well-defined indices according to \cref{thm:magnetization} and \cref{thm:edge_magnetization}. In particular the bulk-edge correspondence holds: $\IB'=\IE'$.
\end{cor}

Thus the composition of stretch function and magnetization or quantized pumping provides indices for any $\UB(\T)$ with mobility gap $\Delta$ and circumvent the relative construction. Note that if $\Delta$ is a spectral gap then $\VB(\T) = \Id$ so that $\IB'=W(\VB)$ and $\IE'$ coincides with the edge index definition of \cite{SadelSchulz-Baldes17} where a particular stretch function was used.

The proof of \cref{cor:BEC_stretched_operators} is not straightforward as one has to check that the underlying assumptions of \cref{thm:magnetization} and \cref{thm:edge_magnetization} are satisfied for $\VB$ and $\VE$, namely that all the properties of $\UB$ and $\UE$ are correctly transfered through the stretch-function construction. This is done in \cref{sec:proof_cor_BECstretched}. 

It is finally legitimate to ask if the two constructions coincide since the relative indices, $\IB$ and $\IE$, and the ones defined through stretch functions, $\IB'$ and $\IE'$, are both defined in a general mobility gap situation. 

\begin{thm}\label{thm:F_invariance}
	If $\UB(\T)$ has a mobility gap $\Delta$ and $F_\Delta$ is a stretch function, then
	\begin{align}
	\IB' &= \IB.
	\end{align}
	In particular $\IB'$ is independent of the choice of stretch function. Moreover by the respective bulk-edge correspondences one infers $\IE'=\IE$.
\end{thm}

Although this last result is to be expected, its poof is actually not straightforward. By \cref{cor:BEC_stretched_operators} we know that $\IB'$ coincides with the relative construction applied to $V$, namely $\MMB(V) = W(\VB^\mathrm{rel})$. But in order to show that $W(\VB^\mathrm{rel})=W(\UB^\mathrm{rel}) \equiv \IB$ we use a smooth deformation of the stretch function from $F_\Delta$ to the identity. Then we have to show that $W$ stays continuous under this deformation. The only other deterministic proof of continuity for indices in the mobility gap regime so far was in \cite{EGS_2005} for the deformation corresponding to tuning the Fermi energy $E_F$ within the mobility gap. Thus the proof of \cref{thm:F_invariance} provides another continuity proof for the index $W$ along a different path and paves the way for further development of locally constant indices at strong disorder.

\begin{rem}It is worth pointing out that it is \cref{thm:F_invariance} which proves that $\mathcal{I}$ is independent of the choice of branch cut $\lambda\in\Delta$ (part of item (a) of \cref{thm:BEC_MG}), since $\mathcal{I}'$ is manifestly independent of $\lambda$.
\end{rem}

\section{Bulk-edge correspondence for the relative evolution \label{sec:RelativeBEC}}

The bulk-edge correspondence was established in \cite{GrafTauber18} in the case where $\UB(\T)$ has a spectral gap. In that case all the operators involved are local in the sense of \cref{assu:locality of Hamiltonian}. In particular $\UB$, $\HE$ and $\UE$ are local, uniformly in $t \in [0,\T]$ (see \cite[Proposition 4.7]{GrafTauber18}). These properties are \emph{independent} of the existence (any) gap of $\UB(\T)$ since they probe the dynamics only in a compact time interval, and hence, at a finite distance from the spectrum, so they remain true also in our setting.

Furthermore when the branch cut of the logarithm is taken inside a spectral gap, $H_\lambda$ (and thus $U_\lambda$) are also local. This is not the case anymore in the mobility gap regime. However the logarithm still has some off-diagonal decay properties that suffice to generalize the proof of the bulk-edge correspondence in the relative construction, as we shall now show.
\subsection{The weakly-local star-algebra}\label{sec:the-weakly-local-star-algebra}

\begin{defn}\label{def:weakly-local-operator} (\emph{Weakly-local operators}) The operator $A\in\BLO$ is said to be \emph{weakly-local} iff there is some $\nu\geq0$ such that for any $\mu>0$ sufficiently large there is some constant $C_{\mu}<\infty$ with
	\begin{align}\label{eq:weakly-local-condition}
	\|A_{xy}\| \leq C_{\mu} (1+\|x-y\|)^{-\mu}(1+\|x\|)^\nu\qquad(x,y\in\mathbb{Z}^d)\,.
	\end{align}
\end{defn}
In our application, the sufficiently large value of $\mu$ will usually be $2$ (\cref{cor:smooth_f_is_weakly_local}) and fixed throughout for all operators. However, to discuss the algebraic properties we allow this value to be arbitrary.
\begin{rem}
	In the above definition, when $\nu=0$, we call the operator polynomially-local, as opposed to the condition in \cref{assu:locality of Hamiltonian} which is exponentially local or just local. 
	
	In summary, the "weak" qualifier refers to the possible diagonal blowup, which is a consequence of deterministic conditions of localization, whereas the "polynomial" qualifier refers to having applied the smooth functional calculus on any (localized or not) local operator, see \cref{cor:smooth_f_is_weakly_local}. 
	
	This form of any-power-polynomial decay that comes from the smooth functional calculus is associated with probing the system at infinite times in the weakest possible way (cf. the Borel bounded functional calculus which probes this directly, and only localization could guarantee that it then takes local operators to (exponentially) weakly-local operators).
\end{rem}
\begin{rem}
	Clearly \cref{def:mobility gap} entails that the functional calculus of $\UB(1)$ is weakly-local uniformly as functions vary in $B_1(\Delta)$.
\end{rem}

\begin{rem}
	One could choose various other ways to encode the off-diagonal decay of an operator. Compare with \cite[Section 3.3]{EGS_2005}, which illustrates how to encode (exponential) decay either with bounds on matrix elements or by estimates on the operator norm of a space-weighted version of the operator. Here we refrain from reformulating \cref{eq:weakly-local-condition} in different ways.
\end{rem}

\begin{lem}\label{lem:transpose-preserves-weakly-local-property}
	The transpose of a weakly-local operator is again weakly-local.
	\begin{proof}
		Assume $A$ obeys \cref{def:weakly-local-operator}. Then picking $\mu > \nu$, \begin{align*}
			\|A_{xy}\| &\leq C_\mu (1+\|x-y\|)^{-\mu} (1+\|x\|)^\nu \\
			& \leq C_\mu (1+\|x-y\|)^{-(\mu-\nu)} (1+\|y\|)^\nu(1+\|y\|)^{-\nu}(1+\|x-y\|)^{-\nu}(1+\|x\|)^\nu\,.
		\end{align*}
		But now, $
			\frac{1+\|x\|}{(1+\|y\|)(1+\|x-y\|)} \leq \frac{1+\|x\|}{1+\|y\|+\|x-y\|} 
		$ and using the reverse triangle inequality, $\|x-y\|\geq\|x\|-\|y\|$ so that this fraction is smaller than or equal to one. So is its $\nu$th power.
		
		We find that $\|A_{xy}\| \leq C_{\mu+\nu} (1+\|y-x\|)^\mu (1+\|y\|)^\nu$ for all $\mu$ sufficiently large; in other words, $A^T$ is weakly-local (though with different constants).
		
	\end{proof}
\end{lem}

\begin{lem}\label{lem:weakly-local-operators-form-algebra}
	The weakly-local operators form a star-algebra.
	\begin{proof}
		Due to \cref{lem:transpose-preserves-weakly-local-property} the linearity of taking matrix elements and the triangle inequality of the matrix norm, we only verify the product property. Let $A,B$ be two given weakly-local operators with constants $C_\mu^A,C_\mu^B$ respectively. Then for any $\mu>0$ sufficiently large (for both $A$ and $B$) and $\nu:=\max(\{\nu_A,\nu_B\})$ we have,
		\begin{align}
			\|(AB)_{xy}\| & \leq \sum_z \|A_{xz}\| \|B_{zy}\|\nonumber \\
						  &	\leq \sum_z C^A_\mu (1+\|x-z\|)^{-\mu} (1+\|x\|)^\nu C^B_{\mu} (1+\|z-y\|)^{-\mu}(1+\|z\|)^\nu \label{eq:product-of-weakly-local-is-weakly-local}\\
						  & \leq C^A_\mu C^B_{\mu} (1+\|x\|)^\nu \sum_z  (1+\|x-z\|)^{-\mu}   (1+\|z-y\|)^{-\mu}(1+\|z\|)^\nu\,.\nonumber					  
		\end{align}
		Now note that $(1+\|x-z\|)(1+\|z-y\|) \geq 1 + \|x-z\| + \|z-y\| \geq 1 + \|x-y\|$ so that \begin{align*}
			\|(AB)_{xy}\| & \leq C^A_\mu C^B_{\mu} (1+\|x\|)^\nu (1+\|x-y\|)^{-\mu/2}\\ & \hspace{3cm}\times \sum_z  (1+\|x-z\|)^{-\mu/2}   (1+\|z-y\|)^{-\mu/2} (1+\|z\|)^\nu \,.
		\end{align*}
		Assume further that $\mu > 2\nu$ has been chosen. Then $(1+\|x-z\|)^{-\mu/2}(1+\|z\|)^\nu \leq (\frac{1+\|z\|}{1+\|x-z\|})^{\nu}\leq(1+\|x\|)^\nu$ by $(1+\|x\|)(1+\|x-z\|)\geq1+\|x\|+\|x-z\|\geq1+\|z\|$. We conclude that \begin{align*}
			\|(AB)_{xy}\| & \leq C^A_\mu C^B_{\mu} (1+\|x\|)^{2\nu} (1+\|x-y\|)^{-\mu/2}\sum_z     (1+\|z-y\|)^{-\mu/2} \\
			& \leq C^A_\mu C^B_{\mu} \bigl(\sum_{z\in\mathbb{Z}^d}(1+\|z\|)^{-\mu/2}\bigr)  (1+\|x-y\|)^{-\mu/2}(1+\|x\|)^{2\nu}\,.
		\end{align*}
		If now we also pick $\mu$ large enough so that the sum in the first parenthesis is finite (e.g. $\mu>2(d+1)$) then we find our result.
	\end{proof}
\end{lem}
\subsection{The logarithm is weakly-local}
For the rest of this section we assume that there is some non-empty interval $\Delta\subseteq\SO$ which is a mobility-gap for $\UB(\T)$ in the sense of \cref{def:mobility gap}. We further assume that $\lambda\in\Delta$, where $\lambda$ is the position of the branch cut used in the definition of $H_\lambda$ from \cref{eq:effective Hamiltonian}.

\begin{lem}\label{lem:extend mobility gap condition to smooth maps with jump discontinuities within the gap}
	$f(\UB(1))$ is also weakly-local for all bounded $f:\SO\to\mathbb{C}$ which are smooth outside of $\Delta$ and piecewise smooth with a finite number of jump discontinuities within $\Delta$.
	\begin{proof}
		Assume that $f$ has \emph{one} jump discontinuity at some $\lambda_0\in\Delta$ and is otherwise smooth. Since $\Delta$ is an interval, pick some other $\lambda_1\in\Delta\setminus\{\lambda_0\}$. For each $\Omega\in\{(\lambda_0,\lambda_1),\SO\setminus(\lambda_0,\lambda_1)\}=:S$, the restriction $\left.f\right|_{\Omega}:\Omega\to\mathbb{C}$ is smooth and so has a smooth extension $f_\Omega^s:\SO\to\mathbb{C}$ (that is $\left.f\right|_{\Omega}=\left.f_\Omega^s\right|_{\Omega}$). Hence $\chi_\Omega f = \chi_\Omega f_\Omega^s$ and $f = \sum_{\Omega\in S}f\chi_\Omega = \sum_{\Omega\in S}f_\Omega^s \chi_\Omega$. 
		Any smooth function of a local unitary operator is also weakly-local by \cref{cor:smooth_f_is_weakly_local}.
		On the other hand, $\chi_\Omega\in B_1(\Delta)$, so the corresponding operator is weakly-local by the assumption entailed in \cref{def:mobility gap}. Thus \cref{lem:weakly-local-operators-form-algebra} allows us to conclude about the whole of $f$. 
		
 	\end{proof}
\end{lem}

\begin{cor}\label{cor:effective Hamiltonian and its evolution are weakly-local}
	Both $H_\lambda$ and $U_\lambda$ are weakly-local.
	\begin{proof}
		Since $\lambda\in\Delta$, we get that $\log_\lambda$ is analytic except for a jump discontinuity within $\Delta$ as $f$ of \cref{lem:extend mobility gap condition to smooth maps with jump discontinuities within the gap}. Now $U_\lambda(t) = (\ee^{-\ii t\, \cdot} \circ \ii \log_\lambda)(\UB(1)) $, $\ee^{-\ii t \, \cdot}$ is analytic, so that for fixed $t$, the composition $\ee^{-\ii t\, \cdot} \circ \ii \log_\lambda$ is again analytic apart from one jump discontinuity within $\Delta$, which is covered by \cref{lem:extend mobility gap condition to smooth maps with jump discontinuities within the gap}.
	\end{proof}
\end{cor}

\subsection{The weakly-local-and-confined two-sided ideal}\label{subsec:the-weakly-local-and-confined-two-sided-ideal}

\begin{defn}\label{defn:weakly-local-and-confined-operator} (\emph{Weakly-Local-and-Confined Operators}) The operator $A\in\BLO$ is said to be \emph{weakly-local-and-confined in direction $i$} for $i=1,\dots,d$ iff there is some $\nu>0$ such that for any $\mu>0$ sufficiently large there is some constant $C_\mu<\infty$ with
	\begin{align}\label{eq:weakly-local-confined-condition}
	\|A_{xy}\| \leq C_{\mu} (1+\|x-y\|)^{-\mu}(1+|x_i|)^{-\mu}(1+\|x\|)^\nu\qquad(x,y\in\mathbb{Z}^d)
\end{align}
We see that adding the "confined" condition guarantees that the operator has also diagonal decay at least in one direction.
\end{defn}
\begin{lem}\label{lem:weakly-local-and-confined-symmetric-in-x-y}
	If $A$ is weakly-local and confined in direction $i$ then so is $A^T$.
	\begin{proof}
		Since $\|x\|\geq|x_i|$, we have $(1+\|x\|)^{-\mu}\leq(1+|x_i|)^{-\mu}$, and hence, \begin{align*}
			\|A_{xy}\| &\leq C_{\mu} (1+\|x-y\|)^{-\mu/2}(1+|x_i-y_i|)^{-\mu/2}(1+|x_i|)^{-\mu}(1+\|x\|)^\nu \\
			&\leq C_{\mu} (1+\|x-y\|)^{-\mu/2}((1+|x_i-y_i|)(1+|x_i|))^{-\mu/2}(1+\|x\|)^\nu
		\end{align*}
		Now, $(1+|x_i-y_i|)(1+|x_i|) \geq 1+|x_i-y_i|+|x_i| \geq 1+|y_i|$, so that 
		\begin{align*}
			\|A_{xy}\| \leq C_{\mu} (1+\|x-y\|)^{-\mu/2}(1+|y_i|)^{-\mu/2}(1+\|x\|)^\nu
		\end{align*}
		Now we can follow the same procedure as in \cref{lem:transpose-preserves-weakly-local-property} to replace the $(1+\|x\|)^\nu$ factor with a $(1+\|y\|)^\nu$ (by worsening the constants).
	\end{proof}
\end{lem}

\begin{lem}\label{lem:operator-norm-characterization-of-weakly-local-confined-operators}
	If $A$ is weakly-local-and-confined in direction $i$, then for all $\mu$ sufficiently large and $\nu$ as in \cref{defn:weakly-local-and-confined-operator} we have $\|(1+\|X\|)^{-\nu}(1+|X_i|)^\mu A\|<\infty$.
	\begin{proof}
		We use Holmgren's bound and the assumed bound in \cref{defn:weakly-local-and-confined-operator} to get \begin{align*}
			\|(1+\|X\|)^{-\nu}(1+|X_i|)^\mu A\| &\leq \max_{x\leftrightarrow y}\sup_y \sum_x \|((1+\|X\|)^{-\nu}(1+|X_i|)^\mu A)_{xy}\| \\
			&\leq \max_{x\leftrightarrow y}\sup_y \sum_x (1+\|x\|)^{-\nu}(1+|x_i|)^\mu\|A_{xy}\| \\
			& \leq \max_{x\leftrightarrow y}\sup_y \sum_x (1+\|x\|)^{-\nu}(1+|x_i|)^\mu \\
			& \hspace{4cm} C_{\mu} (1+\|x-y\|)^{-\mu}(1+|x_i|)^{-\mu}(1+\|x\|)^\nu \\
			& = C_\mu \sum_{x\in\mathbb{Z}^d} (1+\|x\|)^{-\mu}<\infty\,,
		\end{align*}
		assuming $\mu$ is chosen sufficiently large so that this last sum is finite.
	\end{proof}
\end{lem}

\begin{lem}\label{lem:algebraic-properties-of-weakly-local-confined-operators}
	The space of weakly-local-and-confined in direction $i$ operators forms a star-closed two-sided ideal within the star-algebra of weakly-local operators.
	\begin{proof}
		The additive subgroup property follows by the linearity of taking matrix elements as well as the triangle inequality of the matrix norm associated to $\mathbb{C}^N$. The star-closure follows due to \cref{lem:weakly-local-and-confined-symmetric-in-x-y}.
		
		Let now $A$ be weakly-local and confined in direction $i$ and $B$ be merely weakly-local. Then pick $\mu>0$ sufficiently large for both $A$ and $B$ and let $\nu:=\max(\{\nu_A,\nu_B\})$, to get \begin{align*}
			\|(AB)_{xy}\| &\leq \sum_z \|A_{xz}\|\|B_{zy}\| \\
			&\leq \sum_z C^A_{\mu} (1+\|x-z\|)^{-\mu}(1+|x_i|)^{-\mu}(1+\|x\|)^{\nu_A} C^B_\mu (1+\|z-y\|)^{-\mu}(1+\|z\|)^{\nu_B} \\
			&= C^A_{\mu}C^B_\mu(1+|x_i|)^{-\mu}\sum_z (1+\|x-z\|)^{-\mu}(1+\|x\|)^{\nu}  (1+\|z-y\|)^{-\mu}(1+\|z\|)^{\nu}\,.
		\end{align*}
		Everything after $(1+|x_i|)^{-\mu}$ is identical to \cref{eq:product-of-weakly-local-is-weakly-local} (after which we showed that the remainder expression is estimated as weakly-local), so that we find $AB$ is also weakly-local and confined in direction $i$.
		
		Since $BA = (A^\ast B^\ast)^\ast$, $A^\ast$ is weakly-local and confined in direction $i$, $B^\ast$ is weakly-local, so that by the previous paragraph, $A^\ast B^\ast$ belongs to this ideal as well, and hence by the star-closure, $BA$ as well.
	\end{proof}
\end{lem}
\begin{lem} If $A,B$ are weakly-local-and-confined in direction $i,j$ respectively, then $AB$ is weakly-local and confined in directions $i$ and $j$ simultaneously.
	\begin{proof}
		Due to \cref{lem:weakly-local-and-confined-symmetric-in-x-y} we may interchange which of the indices of the matrix element we want to represent the confinement. Thus we are allowed to write, for $\mu>0$ sufficiently large for both $A$ and $B$ and $\nu:=\max(\{\nu_A,\nu_B\})$ \begin{align*}
		&\|(AB)_{xy}\| \\ &\leq \sum_z \|A_{xz}\|\|B_{zy}\| \\
		&\leq \sum_z C^A_{\mu} (1+\|x-z\|)^{-\mu}(1+|x_i|)^{-\mu}(1+\|x\|)^{\nu} C^B_\mu (1+\|z-y\|)^{-\mu}(1+|y_j|)^{-\mu}(1+\|z\|)^{\nu} \\
		&\leq C^A_\mu C^B_\mu (1+|x_i|)^{-\mu}(1+|y_j|)^{-\mu}\sum_z  (1+\|x-z\|)^{-\mu}(1+\|x\|)^{\nu}(1+\|z-y\|)^{-\mu}(1+\|z\|)^{\nu}\,.
		\end{align*}
		Now, by \cref{lem:weakly-local-operators-form-algebra} we know that the expression from $\sum_z$ and after is estimated by something which is weakly-local. Then we may again use \cref{lem:weakly-local-and-confined-symmetric-in-x-y} to replace the $(1+|y_j|)^{-\mu}$ factor with $(1+|x_j|)^{-\mu}$.
	\end{proof}
\end{lem}

As in \cite{EGS_2005}, multiplying $d$ weakly-local-and-confined operators (each in a distinct direction of all possible directions in $\mathbb{Z}^d$)  gives trace class operators. Here our notion of confined is however weaker because we have merely polynomial decay, which changes very little. We denote the trace norm by $\norm{\cdot}_1$.

Since this paper uses $d=2$ that's the scope of the lemma below, whose generalization to arbitrary $d$ is straight-forward.

\begin{lem}\label{lem:product_of_confined_is_trace_class}
	If $A,B$ are both weakly-local, $A$ also confined in direction $1$ and $B$ in direction $2$, then $\|AB\|_1<\infty$.
	\begin{proof}
		Assume $\mu>0$ is sufficiently large for both $A$ and $B$ and let $\nu:=\max(\{\nu_A,\nu_B\})$. Using the freedom that \cref{lem:weakly-local-and-confined-symmetric-in-x-y} affords, we may estimate
		\begin{align*}
			& \|AB\|_1 \\ &\leq \sum_{xyz} \|A_{xz}\|\|B_{zy}\| \\
			& \leq \sum_{xyz} C^A_\mu (1+\|x-z\|)^{-\mu}(1+|z_1|)^{-\mu}(1+\|z\|)^{\nu}C^B_\mu (1+\|z-y\|)^{-\mu}(1+|z_2|)^{-\mu}(1+\|z\|)^{\nu}\,.
		\end{align*}
		Now, $(1+|z_1|)(1+|z_2|)\geq1+|z_1|+|z_2|\equiv1+\|z\|$ so that we find, by summing first over $z$ and then using translation invariance for the $x$ and $y$ sums, \begin{align*}
			\|AB\|_1 &\leq (\sum_x (1+\|x\|)^{-\mu})(\sum_y (1+\|y\|)^{-\mu})(\sum_z (1+\|z\|)^{-(\mu-2\nu)})\,.
		\end{align*}
		If we pick $\mu>0$ sufficiently large so that all three sums are finite (e.g. $\mu\geq 2\nu+d+1$) then $AB$ is indeed trace-class.
	\end{proof}
\end{lem}

\begin{lem}\label{lem:estimate on difference of switch functions}
	From \cite[Proof of Lemma 2]{Graf_Shapiro_2018_1D_Chiral_BEC} we use: for any switch function $\Lambda:\mathbb{R}\to\mathbb{R}$ we have the estimate: for any $\mu>0$ we have some $C_{\Lambda\mu}<\infty$ such that $|\Lambda(n)-\Lambda(m)|\leq C_{\Lambda\mu}(1+|n-m|)^{+\mu}(1+\frac{1}{2}|n|)^{-\mu}$ for all $n,m\in\mathbb{Z}$.
\end{lem}

\begin{cor}\label{cor:switch-function-makes-weakly-local-into-confined}
	If $A$ is weakly-local then $\partial_i A$ is weakly-local and confined in direction $i$.
	\begin{proof}
		We have by the previous estimate on $\Lambda$, for any $\mu'>0$ and $\mu>0$ sufficiently large for $A$, \begin{align*}
			\|(\partial_i A)_{xy}\| & = |\Lambda(x_i)-\Lambda(y_i)|\|A_{xy}\| \\
			& \leq C_{\Lambda\mu'}(1+|x_i-y_i|)^{+\mu'}(1+\frac{1}{2}|x_i|)^{-\mu'} C_{A\mu} (1+\|x-y\|)^{-\mu}(1+\|x\|)^\nu \\
			&\leq  C_{\Lambda\mu'}C_{A\mu}(1+\|x-y\|)^{-(\mu-\mu')}(1+\frac{1}{2}|x_i|)^{-\mu'}  (1+\|x\|)^\nu \,.
		\end{align*}
		By adjusting the constants we can remove the factor $\frac{1}{2}$, and then always taking the worst rate of decay we find the form of \cref{eq:weakly-local-confined-condition}.
	\end{proof}
\end{cor}

\begin{cor}\label{cor:switch_to_traceclass}
	As in \cite[Lemma 2]{Graf_Shapiro_2018_1D_Chiral_BEC} for one-dimension, we have in two-dimensions: if $A$ is weakly-local then $\|(\partial_1 A) \partial_2 A\|_1<\infty$. Moreover if $A$ is also confined in direction $i$ then $\|\partial_j A \|_1<\infty$ for $j \neq i$.
\end{cor}

\subsection{The bulk-edge correspondence in the relative construction}\label{subsec:The relative BEC}
The central ingredient of the bulk-edge correspondence is the relation between truncated bulk and edge propagator. For $\HB$ local, the difference $D = \UE-\widehat{\UB}$ is local and confined in direction 1, uniformly in $t \in [0,\T]$. This result is also independent of the existence of (any) gap of $\UB(1)$ (see \cite[Proposition 4.10]{GrafTauber18}), and is generalized from local to weakly-local operators below.

\begin{lem}\label{lem:definition_of_D} Let $H^\circ:\SO\to\BLO$ be some weakly-local Floquet Hamiltonian, with its associated $H^\circ_{\mathrm{E}}$, $\UB^\circ$ and $\UE^\circ$ also weakly-local. Then $D^\circ := \UE^\circ-\widehat{\UB^\circ}$ is weakly local and confined in direction $1$, uniformly in $t \in [0,\T]$.
	
\begin{proof} 
	To deal with $D^\circ$ we recall that $\widehat{\UB^\circ} \equiv \iota^\ast \UB^\circ \iota$.
	Since $\UE^\circ$ is weakly-local, by \cref{lem:weakly-local-operators-form-algebra} it suffices to deal with $\Id-\iota^\ast \UB^\circ(t) \iota \UE^\circ(t)^\ast= \int_{s=0}^t \partial_s(\iota^\ast \UB^\circ(s) \iota \UE^\circ(s)^\ast) \dd{s} $. Since all estimates about the weak-locality of the involved operators are uniform in time (the time interval is compact), the weak-locality and confined property of the integrand implies that of the integral. With the shorthand notation $\partial$ for derivative w.r.t. time and using the defining property $\ii \partial U = H U$ and the adjoint of this equation, and finally the fact that a Hamiltonian and the semi-group which it generates commute, we have, \begin{align*}
			\partial(\iota^\ast \UB^\circ \iota (\UE^\circ)^\ast) &= \iota^\ast (\partial \UB^\circ) \iota (\UE^\circ)^\ast + \iota^\ast \UB^\circ \iota \partial (\UE^\circ)^\ast \\
			& = \iota^\ast (-\ii \HB^\circ \UB^\circ) \iota (\UE^\circ)^\ast + \iota^\ast \UB^\circ \iota (+\ii (\UE^\circ)^\ast \iota^\ast H_{\lambda} \iota) \\
			& = -\ii \iota^\ast \UB^\circ \HB^\circ  \iota (\UE^\circ)^\ast + \ii \iota^\ast \UB^\circ \iota \iota^\ast H_{\lambda} \iota (\UE^\circ)^\ast  \\
			& = \ii \iota^\ast \UB^\circ (\iota \iota^\ast-\Id) \HB^\circ  \iota (\UE^\circ)^\ast \,.
		\end{align*}
		We note that $\iota \iota^\ast-\Id = \Theta_1^\perp$ where $\Theta$ is the step-function--a valid choice of switch function. In fact all that matters now is that we found a factor $\Lambda_1^\perp \HB^\circ \iota$, and then, using $\Lambda_1^\perp \iota = 0$, this factor equals $[\Lambda_1^\perp, \HB^\circ]\iota = \ii (\partial_1 \HB^\circ) \iota$. But $\HB^\circ$ is weakly-local and we now invoke \cref{cor:switch-function-makes-weakly-local-into-confined} to assert its spatial derivative is confined.
	\end{proof}
\end{lem}

In particular this Lemma applies to $H^\circ = H_\lambda$ and $H^\circ = \HB^\mathrm{rel}$.

\begin{proof}[Proof of \cref{thm:BEC_MG}] 
One adapts \cite{GrafTauber18} from local to weakly-local operators. The first step is to show that the indices are well-defined, the second that they coincide. The strategy is the following: when all the trace quantities involved are finite, all the other computations follow \cite{GrafTauber18}. 

By \cref{lem:definition_of_D} applied to $H^\mathrm{rel}$ we deduce that $\UE^\mathrm{rel}(\T) = 1 + D^\mathrm{rel}(\T)$, and consequently $[\Lambda_2,\UE^\mathrm{rel}(\T)] = -\ii \partial_2 D^\mathrm{rel}(\T)$ which is trace class by \cref{cor:switch_to_traceclass}, so that $\IE$ is well defined. The invariance under the choice of switch function is a simple computation, and the facts it is integer valued is achieved by choosing $\Lambda_2 = \Theta(X_2)$ and identifying $\IE$ with an index of a pair of projections. See \cite[Proposition 3.2]{GrafTauber18} for more details. Then by  \cref{lem:algebraic-properties-of-weakly-local-confined-operators}, \cref{cor:switch-function-makes-weakly-local-into-confined} and \cref{lem:product_of_confined_is_trace_class},  $[\Lambda_i, \UB^\mathrm{rel}](\UB^\mathrm{rel})^*[\Lambda_j, \UB^\mathrm{rel}] = -\partial_i (\UB^\mathrm{rel}) (\UB^\mathrm{rel})^*\partial_j \UB^\mathrm{rel}$ is trace-class for all $t \in [0,\T]$ and $i\neq j$ so that $\IB=W(\UB^\mathrm{rel})$ is finite. Similarly, the switch-function independence and the integer value follow from \cite[Proposition 3.3]{GrafTauber18}.

The proof of the bulk-edge correspondence is now reduced to an algebraic computation that involve trace class operators. Passing from local to weakly-local operators does not modify the finite trace of the expression, mostly because the switch function formalism works the same, as we have already seen in \cref{lem:definition_of_D}. We first rewrite $\IE=\lim_{r \rightarrow \infty} \tr (\UE^\mathrm{rel}(\T)^\ast[\Lambda_2,\UE^\mathrm{rel}(\T)] Q_{1,r})$ where $Q_{1,r} = \Theta(r-X_1)$ is a cut-off in direction 1 on $\HHE$ for $r \in \mathbb N$. At finite $r$ the previous expression becomes trace-class for every $t\in [0,\T]$ so we rewrite it as the integral of its derivative. After some algebra we end up with
\begin{align}\label{eq:algebraicBEC}
\tr (\UE^\mathrm{rel}(\T)^\ast[\Lambda_2,\UE^\mathrm{rel}(\T)] Q_{1,r}) = W_r(\UB^\mathrm{rel}) + o(r)
\end{align}
where $W_r$ is given by \eqref{eq:Cartan-Mauerer-form} for $\Lambda_1 = 1-Q_{1,r}$. Since this quantity is independent of the choice of switch function we conclude $W_r(\UB^\mathrm{rel}) = W(\UB^\mathrm{rel})$ so that $\IE = \IB$ in the $r\rightarrow \infty$ limit. Equality \eqref{eq:algebraicBEC} only uses $\HE=\widehat \HB$, \cref{lem:definition_of_D} and some updated version of it, namely that 
\begin{align}
[\Lambda_2,\UE^\mathrm{rel}(t)]\UE^\mathrm{rel}(t)^\ast-\iota^*\, [\Lambda_2,\UB^\mathrm{rel}(t)]\UB^\mathrm{rel}(t)^\ast \, \iota
\end{align}
is trace class for $t \in [0,t]$ (see \cite[Lemma 5.5]{GrafTauber18}). All the rest follows by simple algebraic manipulations and the fact that $\tr(A B_n) \rightarrow \tr(A B)$ for $A$ trace class and $B_n \rightarrow B$ strongly. See \cite[Theorem 3.4]{GrafTauber18} for more details.

\end{proof}
We note that, surprisingly, identity \eqref{eq:algebraicBEC} remains true when applied to $\UB$ and $\UE$ instead of the relative evolutions, even if $\UB(\T) \neq \Id$. One still has $W_r(\UB) = W(\UB)$ but this quantity is not quantized anymore. Yet the left hand side of \eqref{eq:algebraicBEC} is converging to it in the $r\rightarrow \infty$ limit but does not coincide with any edge index because the $[\Lambda_2,\UE(\T)]$ is not anymore trace-class. Although not relevant here, this identity will be used below.

\section{The completely localized case \label{sec:completely_localized}}
This section is dedicated to the proof of \cref{thm:magnetization} and \cref{thm:edge_magnetization}. We start by studying the bulk part $\MMB$. Let us assume that $\UB$ is completely localized in the sense of \cref{def:completely localized U}, and $\lambda \in S^1 = \Delta$. According to \cref{thm:BEC_MG}, the bulk index $\IB = W(\UB^\mathrm{rel})$ is well defined. By \eqref{time_concatenation} and \eqref{eq:Cartan-Mauerer-form} we deduce
\begin{align*}
	\IB &= W(\UB) - W(U_\lambda)
\end{align*}
where $U_\lambda(t) \equiv \ee^{-\ii H_\lambda t}$. Since both $\UB$ and $U_\lambda$ are weakly-local, $W(\UB)$ and $W(U_\lambda)$ are finite, although they are not separately integer-valued. We rewrite, for any weakly-local $V:[0,1] \rightarrow \mathcal B(\mathcal H)$ and its generator $H_V := \ii \dot{V} V^\ast $
\begin{align*}
W(V) & = \dfrac{\ii}{2} \int_0^\T \tr \eab H_V [\Lambda_\alpha, V] [\Lambda_\beta, V^\ast]  
\end{align*}
where we have used $[\Lambda, V^\ast] = - V^\ast [\Lambda, V] V^\ast$ and defined $\eab$ anti-symmetric with $\ve_{12} = 1$ (summation over $\alpha,\beta \in {1,2}$ is understood when indices appear twice). Since $\tr [\Lambda_\alpha, H_V V [\Lambda_\beta, V^\ast]] = 0$ and $\eab H_V V [\Lambda_\alpha, [\Lambda_\beta, V^\ast]]=0$ we deduce
\begin{align*}
W(V) & = \dfrac{\ii}{2} \int_0^\T \tr \eab [\Lambda_\alpha, H_V] V [\Lambda_\beta, V^\ast] \,.
\end{align*}
Defining $\delta_\alpha^V := V^\ast [\Lambda_\alpha, V]$ and noticing that 
\begin{align}\label{delta_alpha_dot}
\dot{\delta}_\alpha^V = - \ii V^\ast [\Lambda_\alpha, H_V] V
\end{align}
we deduce
\begin{align}\label{eq:W_wrt_delta}
W(V) &= \dfrac{1}{2} \int_0^\T \tr \eab \delta_\alpha^V \dot{\delta}_\beta^V.
\end{align}
Consequently, we rewrite
\begin{align*}
\IB &= \dfrac{1}{2} \tr \int_0^\T \eab \big(\delta_\alpha \dot{\delta}_\beta- \delta^\lambda_\alpha \dot{\delta}^\lambda_\beta)
\end{align*}
where we use the shorthand notation $\delta_\alpha$ (resp. $\delta^\lambda_\alpha$) for $\delta_\alpha^\UB$ (resp. $\delta_\alpha^{U_\lambda}$). Note that there is no problem to exchange trace and integral here since both $\eab \delta_\alpha \dot{\delta}_\beta$ and $\eab\delta^\lambda_\alpha \dot{\delta}^\lambda_\beta$ are trace class for all $t \in [0,1]$. Finally 
\begin{align*}
\int_0^\T \eab \big(\delta_\alpha \dot{\delta}_\beta- \delta^\lambda_\alpha \dot{\delta}^\lambda_\beta) &= \int_0^\T \eab \big(\UB^\ast \Lambda_\alpha \UB \dot{\delta}_\beta- U^\ast_\lambda \Lambda_\alpha U_\lambda \dot{\delta}^\lambda_\beta) - \int_0^\T \eab \Lambda_\alpha\big( \dot{\delta}_\beta - \dot{\delta}^\lambda_\beta\big)\,.
\end{align*}
The last term vanishes because it is a total derivative, and by the fact that $\delta_\alpha(0) = \delta^\lambda_\alpha(0) = 0$ and $\delta_\alpha(1) = \delta^\lambda_\alpha(1)$ since $\UB(\T)= U_\lambda(\T)$ (note however that $\delta_\alpha(1) \neq 0$ in general). Hence, by \eqref{delta_alpha_dot} and \eqref{eq:def_MB},
\begin{align*}
\IB &= \tr( M(\UB) - M(U_\lambda))\,.
\end{align*}
This relation is very general and does not require $\UB(\T)$ to be completely localized. However it is equivalent to \eqref{eq:I=M-Mlambda} only in the latter case. Indeed $M(\UB)$ and $M(U_\lambda)$ are not separately trace-class, only their difference is. When $\UB(\T)$ is completely localized, the trace of this difference can be computed through its eigenbasis:
\begin{align*}
\IB &= \sum_{z \in \EVS} \tr P_z \big(M(\UB) - M(U_\lambda) \big) P_z\,.
\end{align*}
What remains to show is that the sum can actually be split into two parts, leading to \eqref{eq:I=M-Mlambda}.

\begin{prop}\label{prop:vanishing_magnetization}
	If $\UB(\T)$ is completely localized in the sense of \cref{def:completely localized U}, then the effective evolution magnetization
	\begin{align*}
	\MMB(U_\lambda) &= \sum_{z \in \EVS} \tr P_z \int_0^1 \Im( U_\lambda^\ast \Lambda_1 H_\lambda \Lambda_2 U_\lambda) P_z
	\end{align*}
	is absolutely convergent and vanishes.
\end{prop}

Thus we are left with $\IB = \MMB(\UB)$, so that $\MMB(\UB)$ is well defined and shares all the properties of $\IB$. This proves the main statement of \cref{thm:magnetization}. In the particular case where $\UB(\T)=\Id$, we have $H_\lambda = 0$ and $\delta_\alpha(\T) =0$, so that $M(\UB)$ is trace-class by the previous computation, and $\IB = W(\UB) = \tr(M(\UB)) = \MMB(\UB)$. Finally the case where $H$ is time-independent is a direct consequence of the latter proposition.

\begin{proof}[Proof of \cref{prop:vanishing_magnetization}]
Since $U_\lambda(t)\equiv\ee^{-\ii t H_\lambda}$ and $U_\lambda(1)=\UB(\T)$ is completely localized, then so are $H_\lambda$ and $U_\lambda(t)$ for $t\in [0,1]$ with the same eigenbasis as $\UB(\T)$. Thus for a fixed $z\in\SO$ one has by functional calculus  
\begin{align}\label{eq:proof_mag_aux}
 P_z \int_0^1 \Im( U_\lambda^\ast \Lambda_1 H_\lambda \Lambda_2 U_\lambda) P_z &=  P_z  \Im(  \Lambda_1 H_\lambda \Lambda_2 ) P_z.
\end{align}

By construction $H_\lambda$ is bounded with a real spectrum that unwinds the circular one of $\UB(1)$ with respect to the branch cut $\lambda$. For each eigenvalue of $\UB(1)$, $z \in \SO$, $r:=\ii \log_\lambda(z) \in \mathbb R$ is an eigenvalue of $H_\lambda$ with same eigenprojection $P_z$. For $x \in \mathbb R$ we define the Fermi projection up to $x$ by $P(x) := \chi_{(-\infty,x)}(H_\lambda)$, so that $P(x) = 0$ for $x < \sigma(H_\lambda)$ and $P(x) = \Id$ for $x \geq \sigma(H_\lambda)$.
We use the following representation of $H_\lambda$ 
\begin{align}
\label{layer_cake}
H_\lambda &= C - \int_{\sigma(H_\lambda)}  P(x) \dd x
\end{align}
where $C = \sup(\sigma(H_\lambda)) \in \mathbb R$. This representation comes from the following functional equality
\begin{align}\label{eq:layer_cake_function}
\int_{\Omega} \chi_{(-\infty,x)}(y) \dd x &= \int_{\Omega} \chi_{(y,\infty)}(x) \dd x = \sup(\Omega) - y
\end{align}
for some interval $\Omega$ and $y \in \Omega$. Inserting \eqref{layer_cake} into \eqref{eq:proof_mag_aux} we get
\begin{align*}
P_z  \Im(  \Lambda_1 H_\lambda \Lambda_2 ) P_z &= - \dfrac{\ii}{2} \int_{\sigma(H_\lambda)} \eab P_z \Lambda_\alpha P(x) \Lambda_\beta P_z \dd x\,.
\end{align*}
Consider $z \in \EVS$ and $x \in \sigma(\HB)$ fixed, and define $P(x)^\perp=\Id-P(x)$. Then either $\ii \log_\lambda(z) > x$, in which case $P_z P(x) = 0$ and $P_z P(x)^\perp = P_z$, or $\ii \log_\lambda(z) \leq x$, in which case $P_z P(x) = P_z$ and $P_z P(x)^\perp = 0$. Therefore
\begin{align}\label{eq:def_Tx}
\dfrac{\ii}{2} \eab P_z \Lambda_\alpha P(x) \Lambda_\beta P_z & = \dfrac{\ii}{2} P_z  \eab P(x)^\perp \Lambda_\alpha P(x) \Lambda_\beta P(x)^\perp P_z - \dfrac{\ii}{2} P_z  \eab P(x) \Lambda_\alpha P(x)^\perp \Lambda_\beta P(x) P_z \cr
& := \dfrac{1}{2} P_z T(x) P_z\,.
\end{align}
Moreover, $T(x)$ is trace-class for every $x \in \sigma(H_\lambda)$. Indeed, after some algebra
\begin{align}\label{eq:Tx_traceclass}
T(x) &= \ii P(x)^\perp\big[[\Lambda_1, P(x)^\perp], [\Lambda_2, P(x)^\perp]\big] - \ii P(x)\big[[\Lambda_1, P(x)], [\Lambda_2, P(x)]\big]
\end{align}
and each term is separately trace-class by \cref{cor:switch_to_traceclass}. Indeed, $P(x) = \chi_{(\lambda,\ee^{-\ii x})}(\UB(\T))$ is weakly-local according to \cref{def:mobility gap} since $\UB(\T)$ is completely localized. Thus for every $z \in \EVS$, $P_z T(x) P_z$ is trace-class (even if $z$ is infinitely degenerate) and moreover 
\begin{align*}
\dfrac{1}{2}\sum_{z \in \SO}\tr P_z T(x) P_z &= \dfrac{1}{2} \tr(T(x) ) = \dfrac{1}{2} (c(P(x))- c(P(x)^\perp) = c(P(x))\,.
\end{align*}
with the sum on the left hand-side that is absolutely convergent. Since $H_\lambda$ is bounded we have
\begin{align*}
\dfrac{1}{2} \int_{\sigma(H_\lambda)} \sum_{z \in \SO } |\tr P_z T(x) P_z | &< \infty\,.
\end{align*} 
By Fubini's theorem and putting everything together, we deduce that $\MMB(U_\lambda)$ is defined by an absolutely convergent sum. Moreover
\begin{align*}
\MMB(U_\lambda) &= - \int_{\sigma(H_\lambda)} c(P(x)) \dd x\,.
\end{align*}
It was shown in \cite[Prop. 2]{EGS_2005} that $c(P_\Omega(H)) = 0$ for any interval $\Omega$ inside the mobility gap of $H$ that contains only finite-multiplicity eigenvalues. Here the entire spectrum of $H_\lambda$ is a mobility gap, but it might contain infinite degenerated eigenvalues $\alpha_1, \ldots, \alpha_M$. Thus for $x \in \sigma(H_\lambda)$ we are left with
\begin{align*}
c(P(x)) &= \sum_{\alpha_i \in (-\infty,x)\cap\sigma(H_\lambda)} c(P_{\alpha_i}) =0
\end{align*}
for a completely localized $\UB(\T)$, see \cref{def:completely localized U}. Thus $\MMB(U_\lambda)=0$.
\end{proof}

\begin{proof}[Proof of \cref{thm:edge_magnetization}]
	Let $n$ be a fixed integer. From \eqref{eq:algebraicBEC} in the proof of \cref{thm:BEC_MG} and the fact that $\UE(\T)^n=\UE(n)$, we have the following identity:
	\begin{align*}
	\lim_{r\rightarrow \infty} \tr \Big( (\UE(\T)^\ast)^n[\Lambda_2, \UE(\T)^n] Q_{1,r} \Big) &= \frac{1}{2}\int_0^n \tr  \dot{\UB} \UB^\ast[[\Lambda_1,\UB]\UB^\ast, [\Lambda_2,\UB]\UB^\ast]\,.
	\end{align*}	

On the left-hand-side, the expression is trace class for every finite $r$  because of cut-off $Q_{1,r}$ and confinement in direction $2$ through $\Lambda_2$. The right-hand-side is expression \cref{eq:Cartan-Mauerer-form} of $W$ but on a time interval $[0,n]$ instead of $[0,\T]$. In partirular it is independent of switch-function $\Lambda_1$, which is why the limit $r \rightarrow \infty$ is finite. If $\UB(n) = \Id$, $W$ would be quantized and define the bulk index, and the limit on the right would be equal to edge index. Nevertheless the previous equation is true for any pair of bulk and edge operators $\UB$ and $\UE$, as long as they are weakly-local and related by \cref{lem:definition_of_D}, although it is not integer-valued. From now we assume $U(1)$ completely localized. Rewriting $W$ as in \eqref{eq:W_wrt_delta} 
\begin{align}\label{eq:aux} 
&\frac{1}{2}\int_0^n \tr  \dot{\UB} \UB^\ast[[\Lambda_1,\UB]\UB^\ast, [\Lambda_2,\UB]\UB^\ast] \cr & = \dfrac{1}{2} \int_0^n \tr \eab \delta_\alpha \dot{\delta}_\beta \cr & = \dfrac{1}{2} \sum_{z \in \mathcal E}  \tr( P_z \int_0^n \eab U^* \Lambda_\alpha U \dot{\delta}_\beta P_z)  - \dfrac{1}{2} \sum_{z \in \mathcal E}  \tr( P_z \int_0^n \eab \Lambda_\alpha \dot{\delta}_\beta P_z)
\end{align}
where $\delta_\alpha = U^* \Lambda_\alpha U - \Lambda_\alpha$. Since $\delta_\alpha \dot{\delta}_\beta$ is trace class, we permute trace and time integral, and then compute this trace in the eigenbasis of $\UB(\T)$. What remains to show is that the two terms in the last formula obtained by splitting $\delta_\alpha$ are separately finite, and then study their $n\rightarrow \infty$ limit. Note that the eigenbasis of $\UB(\T)$ and $\UB(n)$ are the same since $\UB(n) = \UB(\T)^n$, although the eigenvalues are different. The first term in \eqref{eq:aux} is close to magnetisation
\begin{align*}
\dfrac{1}{2} \sum_{z \in \mathcal E}  \tr( P_z \int_0^n \eab U^* \Lambda_\alpha U \dot{\delta}_\beta P_z) &= \sum_{z \in \mathcal E} \tr( P_z \int_0^n \Im(\UB^* \Lambda_1 H \Lambda_2 \UB) P_z) =: \mathcal M_n(\UB)\,.
\end{align*}
Then we use the facts that $\UB(t) = \UB(t-k) \UB(\T)^k$ for $k \leq t < k+1$ and $k \in \{0,\ldots,n-1\}$ and $\UB(\T)^k P_z = z^k P_z$ for $z \in \mathcal E \subset S^1$. Similarly $\UB^*(t) =  (\UB(\T)^*)^k\UB^*(t-k)$ and $(\UB(\T)^*)^k P_z = z^{-k} P_z$. Moreover $H(t+k) = H(t)$. Applying these relations to the previous time integral that we cut into $n$ parts, we get, up to a change of variable
\begin{align*}
\mathcal M_n (\UB) &= n \MMB(\UB)
\end{align*}
so that $\mathcal M_n (\UB)$ is finite and shares all the properties of $\MMB(\UB)$ from \cref{thm:magnetization}. Moreover $n^{-1} \mathcal M_n(\UB) \rightarrow \MMB(\UB)$ when $n \rightarrow \infty$.

The second term of \eqref{eq:aux} is a total derivative and can be simplified to
\begin{align}\label{eq:aux2}
\dfrac{1}{2} \sum_{z \in \mathcal E}  \tr( P_z \int_0^n \eab \Lambda_\alpha \dot{\delta}_\beta P_z) &= \dfrac{1}{2} \sum_{z \in \mathcal E}  \tr( P_z \eab \Lambda_\alpha \UB(n)^* \Lambda_\beta \UB(n) P_z)
\end{align}
since $\delta_\beta(0)=0$ and $\eab \Lambda_\alpha \Lambda_\beta = 0$. Note that $U(n) = U(1)^n = \ee^{-\ii n H_\lambda}$ for $H_\lambda$ defined in \eqref{eq:effective Hamiltonian} and any $\lambda \in S^1 = \Delta$. Then we use the following functional equality for a continuously differentiable $f:[a,b] \rightarrow  \mathbb C$:
\begin{align*}
f(y) &= f(b) - \int_a^b \dd x f'(x) \chi_{(a,x)}(y)
\end{align*}
for $y \in [a,b]$, which is a generalization of \eqref{eq:layer_cake_function}, see also \cite{EGS_2005}. Consequently
\begin{align}\label{eq:U*_integralrepresentation}
\UB(n)^* &= \ee^{\ii n H_\lambda} = \ee^{\ii n b} \Id - \ii n \int_{\sigma(\HB_\lambda)} \ee^{\ii n x} P(x) \dd x
\end{align}
where $P(x) = \chi_{(-\infty,x)}(H_\lambda)$ and $b = \sup(\sigma(H_\lambda))$. When inserting this expression for $\UB(n)^*$ in \eqref{eq:aux2}, the first term vanishes by antisymmetry. In order to show that the second one is finite, we claim that
\begin{align*}
\dfrac{\ii n}{2} \int_{\sigma(H_\lambda)} \dd x \sum_{z \in \mathcal E} \ee^{\ii n x} \tr( P_z \eab \Lambda_\alpha P(x) \Lambda_\beta \UB(n) P_z )
\end{align*}
is absolutely convergent for any fixed $n$. Indeed since $\UB(n)$ commutes with $P_z$ one has
\begin{align*}
P_z  \ii \eab \Lambda_\alpha P(x) \Lambda_\beta \UB(n) P_z = P_z T(x) \UB(n) P_z 
\end{align*}
where $T(x)$ is defined in \eqref{eq:def_Tx}. Moreover $T(x)$ is trace class as pointed out in \eqref{eq:Tx_traceclass} so  the previous sum over $z$ is absolutely convergent for every $x \in \sigma(H_\lambda)$. The integral is then also absolutely convergent because $H_\lambda$ is bounded.

Consequently, \eqref{eq:aux2} can be rewritten as
\begin{align*}
\dfrac{1}{2} \sum_{z \in \mathcal E}  \tr( P_z \eab \Lambda_\alpha \UB(n)^* \Lambda_\beta \UB(n) P_z) = \dfrac{ n}{2} \sum_{z \in \mathcal E} \ee^{-\ii n z} \int_{\sigma(H_\lambda)} \dd x \, \ee^{\ii n x} \tr(P_z T(x) P_z)
\end{align*}
with absolute convergence.
We finally claim that 
\begin{align}\label{aux3}
\lim\limits_{n\rightarrow \infty} \sum_{z \in \mathcal E} \ee^{-\ii n z} \int_{\sigma(H_\lambda)} \dd x \, \ee^{\ii n x} \tr(P_z T(x) P_z) = 0\,.
\end{align}
First for $z\in \mathcal E$ denote  $g_z(x) := \tr(P_z T(x) P_z)$ that is measurable on $\sigma(H_\lambda)$. Then
\begin{align*}
\int_{\sigma(H_\lambda)} \dd x \, \ee^{\ii n x} \tr(P_z T(x) P_z) &:= c_n(g_z) \mathop{\longrightarrow}_{n \rightarrow \infty} 0
\end{align*}
Indeed it is the $n$-th Fourier coefficient of $g_z$ that vanishes in the large $n$ limit by the Riemann-Lebesgue lemma. Finally $\ee^{-\ii n z} c_n(g_z)$ is summable in $z$ and vanishes when $n \rightarrow \infty$ for fixed $z$. Moreover
\begin{align*}
|\ee^{-\ii n z} c_n(g_z)| \leq \int_{\sigma(H_\lambda)} \dd x | \tr (P_z T_x P_z) |
\end{align*} 
is summable in $z$ since $T_z$ is trace-class, leading to \cref{aux3} by dominated convergence, and concluding the proof.
\end{proof} 

\section{The stretch-function construction \label{sec:stretch_function}}

\subsection{Proof of \texorpdfstring{\cref{cor:BEC_stretched_operators}}{Corollary 2.4} \label{sec:proof_cor_BECstretched}}

\cref{cor:BEC_stretched_operators} is a consequence of \cref{thm:magnetization,thm:edge_magnetization}, that both rely on \cref{thm:BEC_MG}, applied to $\VB\equiv F_\Delta(\UB)$ and $\VE\equiv F_\Delta(\UE)$ for a given stretch function $F_\Delta$. By \cref{def:mobility gap}, $\VB(1)$ is completely localized since $F_\Delta \in B_1(\Delta)$, but in order to replace $\UB$ and $\UE$ by $\VB$ and $\VE$ in the previous theorems we need to show that they satisfy all the required properties concerning locality and confinement. We note that \cref{lem:V_is_local,lem:diff_V} below are true regardless of the existence of (any) gap of $\UB(1)$ and moreover all the operators involved are (polynomially) local since $\HB$ is (exponentially) local by \cref{assu:locality of Hamiltonian}, see \cref{cor:smooth_f_is_weakly_local}. 

The existence of the gap only become relevant when we use localization to assert the (weak) locality of the logarithm, which is when we apply \cref{cor:effective Hamiltonian and its evolution are weakly-local} to $V(1)$. When we do that, we actually get expressions like $\log_\lambda \circ F_\Delta$ applied to $U(1)$, which, as in \cref{lem:extend mobility gap condition to smooth maps with jump discontinuities within the gap}, gets decomposed to sums of functions such as $g\circ F_\Delta$ with $g$ smooth, which is a smooth function, or $\chi_{[\lambda,\lambda']} \circ F_\Delta$ which is in $B_1(\Delta)$ and so \cref{def:mobility gap} applies. The conclusion is that the logarithm of $V(1)$ is also weakly-local so that the relative construction could just as well be applied to $V$.

\begin{lem}\label{lem:V_is_local}
	$\VB$ and $\VE$ are (polynomially) local if $\UB$ and $\UE$ are (exponentially) local. Moreover the maps $t \mapsto \VB(t)$ and $t \mapsto \VE(t)$ are strongly differentiable and their respective generators $H_{\VB}= \ii \dot{\VB}\VB^\ast$ and $H_{\VE}= \ii \dot{\VE}\VE^\ast$ are weakly-local.
\begin{proof}
	The first fact is a direct consequence of \cref{cor:smooth_f_is_weakly_local}, $F_\Delta$ being smooth. For the derivatives we compute for $t,s \in [0,\T]$, using \cref{lem:HJ_unitary} and the resolvent identity,
	\begin{align*}
		\VB(s)-\VB(t) &= \dfrac{1}{2\pi\ii} \int \dd z\dd \bar z (\partial_{\bar z} \tilde F_\Delta(z)) R_{\UB(s)}(z) (\UB(s)-\UB(t))R_{\UB(t)}(z)
	\end{align*} 
	where $\tilde F_\Delta$ is a quasi-analytic extension of $F_\Delta$ and $R_{\UB(s)}(z) = (\UB(s)-z)^{-1}$, that is norm-continuous in $s$. Hence
	\begin{align*}
\partial_t \VB(t) &= \mathop{\mathrm{s-lim}}_{s \rightarrow t} \dfrac{\VB(s)-\VB(t)}{s-t} = \dfrac{1}{2\pi\ii} \int \dd z\dd \bar z (\partial_{\bar z} \tilde F_\Delta(z)) R_{\UB(t)}(z) (\partial_t \UB)(t) R_{\UB(t)}(z)\,.
	\end{align*}
Since $\norm{R_{\UB(t)}(z)} \leq C ||z|-1|^{-1}$ and $|\partial_{\bar z} \tilde F| \leq C | |z|-1 |^N$ for some $N \geq 2$ the integral is convergent. Moreover $\HB$ and $\UB$ are local thus so are $\partial_t \UB = -\ii \HB \UB$ and $R_{\UB(t)}$, the latter by the Combes-Thomas estimate. Since $\tilde F_\Delta$ is compactly supported we deduce that $\partial_t V$ is (polynomially) local, and so is $H_{\VB}$ by \cref{lem:weakly-local-operators-form-algebra}. We proceed similarly for $\VE$.
\end{proof}
\end{lem}

\begin{lem}\label{lem:diff_V}
	The differences
	$\VE - \iota^* \VB \iota$ and $H_{\VE}-\iota^* H_{\VB} \iota$
	are weakly-local and confined in direction 1, uniformly in $t \in [0,\T]$.
\begin{proof}
	This looks like a consequence of \cref{lem:definition_of_D} (see also \cite[Prop. 4.10]{GrafTauber18}). However since $\iota^* \UB \iota$ is not a unitary, it is not obvious how to directly implement functional calculus on it. Instead we should first reformulate this result in the bulk picture. In what follows $D$ denotes an operator that is local and confined in direction 1. We claim that
	\begin{align}\label{aux4} 
	\UB = \iota \UE \iota^* + j U_- j^* + D
	\end{align}
	where $j : \mathcal H_- \hookrightarrow \HHB$ and $j^* : \HHB \twoheadrightarrow \mathcal H_-$ with $\mathcal H_- = \ell^2((\mathbb Z \setminus \mathbb N) \times \mathbb Z) \otimes \mathbb C^N$ is the left half space. Note that $jj^\ast = \Id-P_1$, $j^*j = \Id$, and $j^\ast i = \iota^\ast j = 0$. Finally $U_-$ is generated by $H_-:= j^\ast \HB j$, so that both are local like $\HE$ and $\UE$ are. The proof of \eqref{aux4} is completely analogue to \cref{lem:definition_of_D} and relies on the fact that $[P_1,\HB]$ is local and confined in direction 1.
	
	Then we consider the unitary $U_\mathrm{d} := \iota \UE \iota^* + j U_- j^*$, that satisfies $R_{U_\mathrm{d}}(z) =  \iota R_{\UE}(z) \iota^\ast + j R_{U_-}(z) j^*$ where $R_U(z)=(U-z)^{-1}$. By \eqref{aux4} and the resolvent identity we deduce
	\begin{align}\label{aux5}
	R_{\UB}(z) - R_{U_\mathrm{d}}(z) = -R_{\UB}(z) D R_{U_\mathrm{d}}(z)
	\end{align}
	We compute $F_\Delta(\UB)$ and $F_\Delta(U_\mathrm{d})$ through quasi-analytic functional calculus, see \cref{lem:HJ_unitary}, leading to
	\begin{align*}
	F_\Delta(\UB) - F_\Delta(U_\mathrm{d}) = \dfrac{1}{2\pi \ii} \int \dd z \dd \bar z (\partial_{\bar z} \tilde F_\Delta) R_{\UB}(z) D R_{U_\mathrm{d}}(z)\,.
	\end{align*}
	 On the right hand side the integral is convergent because of the decaying behavior of $\partial_{\bar z} \tilde F_\Delta$ around $\SO$, similarly to the previous proof. Moreover both resolvents are local by Combes-Thomas estimate so that the integral is weakly-local and confined in direction 1 by \cref{cor:smooth_f_is_weakly_local}. On the left hand side we have $F_\Delta(U_\mathrm{d}) =  \iota F(\UE) \iota^\ast + j F(U_-) j^*$, so that the difference $\iota^* (F_\Delta(\UB) - F_\Delta(U_\mathrm{d}))\iota = \iota^* V \iota - \VE$ has the expected property.
	
	It is then easy to show that $\partial_t \VE - \iota^* \partial_t V \iota$ is also weakly-local and confined in direction 1, by using quasi-analytic functional calculus of \cref{lem:V_is_local} and the fact that both $\partial_t \UE - \iota^* \partial_t \UB \iota$ and $R_{\UE}(z) - \iota^* R_{\UB}(z) \iota$ are local and 1-confined, respectively coming from \cref{lem:definition_of_D} and \eqref{aux5}. We deduce that $H_{\VE} - \iota^* H_{\VB} \iota$ has the expected property.
\end{proof}
\end{lem}

\subsection{Stretch-function invariance}

\begin{figure}[htb]
	\centering
\begin{tikzpicture}
\newdimen\r
\pgfmathsetlength\r{1.3cm}
\draw[] (0,0) circle (\r);
\draw[densely dashed] (0,0) -- (150:1.3*\r) node[left] {$\lambda$};
\fill[black] (130:\r) circle (0.05);
\node at (120:0.65*\r) {$\mathfrak F_{s_2}(z)$};
\fill[black] (100:\r) circle (0.05);
\node at (90:1.2*\r) {$\mathfrak F_{s_1}(z)$};
\node at (-1.3*\r,-\r) {(a)};
\begin{scope}[xshift=6cm]
\draw[] (0,0) circle (\r);
\draw[densely dashed] (0,0) -- (150:1.3*\r) node[left] {$\lambda$};
\fill[black] (130:\r) circle (0.05);
\node at (120:0.65*\r) {$\mathfrak F_{s_2}(z)$};
\fill[red] (170:\r) circle (0.05);
\node[red] at (185:0.6*\r) {$\mathfrak F_{s_1}(z)$};
\node at (-1.3*\r,-\r) {(b)};
\end{scope}
\end{tikzpicture}
	\caption{In the proof, situation (a) happens as a rule and situation (b) never occurs by choice of $\mathfrak{F}$. \label{fig:stretch-invariance}}
\end{figure}
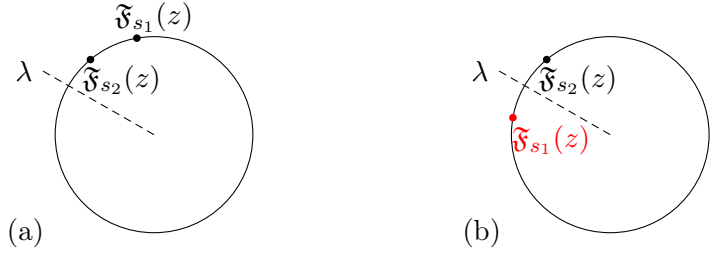

\begin{proof}[Proof of \cref{thm:F_invariance}]

We assume that $F:\mathbb{C}\to\mathbb{C}$ is a stretch function and have $V \equiv F \circ U$. As mentioned we assume $F$ is smooth. Above we have shown that $\IB'=W(V^{\mathrm{rel}})$, so that our task now is to show that $W(V^{\mathrm{rel}})=W(U^{\mathrm{rel}})$. Let $[0,1]\ni s\mapsto \mathfrak{F}_s(\cdot)$ be a homotopy that interpolates smoothly between the identity map $\mathbb{C}\ni z\mapsto z$ at $s=0$ and $F$ at $s=1$. Since $F$ itself is a "stretching" of the mobility gap $\Delta\subseteq\SO$ onto the entire circle, we pick this interpolation such that it stretches \emph{about} the branch cut $\lambda\in\Delta$. This point is crucial and will be used later on, in that it means no eigenvalue of $\mathfrak{F}_s(U(1))$ crosses $\lambda$ as $s$ changes. The gist of the argument is as follows. All maps involved are continuous (even smooth) except one, $\log_\lambda$. While this map indeed has a jump discontinuity, the particular form of deformation which we choose doesn't ever cross this point of discontinuity--in other words, $\lambda$ is a fixed point of the deformation (see \cref{fig:stretch-invariance}).

The smoothness assumption means that, in particular, for fixed $z$, $s\mapsto\mathfrak{F}_s(z)$ is differentiable, for all $s$, $z\mapsto\mathfrak{F}_s(z)$ is smooth (so $\mathfrak{F}_s(U(t))$ is local for all $t$ and it makes sense to take the derivative of $t\mapsto\mathfrak{F}_s(U(t))$) and for fixed $z$, $s\mapsto \mathfrak{F}_s'(z)$ is differentiable. In addition, because $s\mapsto \mathfrak{F}_s$ interpolates between $\Id$ and $F$, the mobility gap never closes (it only gets stretched from $\Delta\to\SO\setminus\Set{1}$) for all $s$, $\lambda$ is within the mobility gap of $\mathfrak{F}_s(U(1))$ so that $\log_\lambda(\mathfrak{F}_s(U(1)))$ is weakly-local for all $s$. 

Hence $W(\mathfrak{F}_s(U)^{\mathrm{rel}})$ is well-defined and integer valued for all $s$, so it suffices to prove that $|W(\mathfrak{F}_{s_1}(U)^{\mathrm{rel}})-W(\mathfrak{F}_{s_2}(U)^{\mathrm{rel}})|<1$ for any $s_1,s_2\in[0,1]$ with $|s_1-s_2|$ sufficiently small. Recall that $W(\mathcal{V}^{\mathrm{rel}}) = W(\mathcal{V}) - W(\mathcal{V}_\lambda)$ so that by the triangle inequality we can work separately with $|W(\mathfrak{F}_{s_1}(U))-W(\mathfrak{F}_{s_2}(U))|$ and $|W(\mathfrak{F}_{s_1}(U)_\lambda)-W(\mathfrak{F}_{s_2}(U)_\lambda)|$, though each part is not separately an integer. To probe the smallness of these expressions we use \cref{lem:W continuity estimate}.

We will use the fact that if $A_n\to A$ strongly and $T$ is a trace-class operator, then $T A_n \to T A$ in trace class norm. This fact is boosted, using the weakly-local properties, to \cref{lem:A_n to A strongly and all weakly local T_alpha confined in alpha} and \cref{lem:A_n to A and B_n to B strongly weakly-local and confined implies trace-class convergence}. Note that in order to use these lemmas, one must have uniform exponents $\mu$ and $\nu$ which is certainly not part of the context in \cref{sec:the-weakly-local-star-algebra}. However, this is actually not a problem since the form of weak-locality that is produced by \cref{cor:smooth_f_is_weakly_local} gives us the ability to choose the minimal exponents $\mu$ once and for all. The exponent $\nu$ is actually not even necessary here since the deformation is always applied on $\UB(1)$ which is honestly local and not just weakly-local, but even if that weren't the case, one can just choose a universal $\nu$ which makes $\sum_x (1+\|x\|)^{-\nu}$ finite.

Since $s\mapsto\mathfrak{F}_s(z)$ is differentiable,  $$\slim_{\ve\to0}\frac{1}{\ve}(\mathfrak{F}_{s+\ve}(U(t)))-\mathfrak{F}_s(U(t))))=\partial_s \mathfrak{F}_s(U(t))),$$so that $\frac{1}{\ve}T( \mathfrak{F}_{s+\ve}(U(t))-\mathfrak{F}_s(U(t))) \to T\partial_s \mathfrak{F}_s(U(t))) $ in trace-class norm for any trace-class $T$. Similarly we handle also $\partial_t\mathfrak{F}_s(U(t))) =  \mathfrak{F}_s'(U(t)))\dot{U}(t)$ which is also differentiable as a function of $s$. Since $\mathfrak{F}_s(U(t)))$ is weakly-local for any value of $s$, we also have similar convergence for the spatial derivatives: $\frac{1}{\ve}T_\alpha\partial_\beta( \mathfrak{F}_{s+\ve}(U(t))-\mathfrak{F}_s(U(t))) \to T_\alpha\partial_\beta \partial_s \mathfrak{F}_s(U(t)))$ in trace-class norm for any $T_\alpha$ which is weakly-local and confined in the $\alpha$ direction. We conclude that $|W(\mathfrak{F}_{s_1}(U))-W(\mathfrak{F}_{s_2}(U))|$ can be made arbitrarily small as $s_2 \to s_1$.

When dealing with $|W(\mathfrak{F}_{s_1}(U))_\lambda)-W(\mathfrak{F}_{s_2}(U)_\lambda)|$, it might appear that we are stuck, since $ \mathfrak{F}_{s_1}(U))^\lambda(t) \equiv \exp(t \log_\lambda(\mathfrak{F}_{s_1}(U(1)))) $ and $\log_\lambda$ is \emph{not} continuous. Furthermore, algebraic laws like $\log(\frac{z_1}{z_2})=\log(z_1)-\log(z_2)$ only hold mod $2 \pi \ii$ in general, which could introduce jump discontinuities. Since $\mathfrak{F}_{s_1}(U(1))$ and $\mathfrak{F}_{s_2}(U(1))$ are functions of the same operator $U(1)$, they commute and hence have the same diagonalization. Indeed, let $P$ be the projection-valued spectral measure of $U(1)$. Then \begin{align*}
\log_\lambda(\mathfrak{F}_{s_1}(U(1))) - \log_\lambda(\mathfrak{F}_{s_2}(U(1))) & = \int_{z\in\SO} \log_\lambda(\mathfrak{F}_{s_1}(z))\dif{P(z)}-\int_{z\in\SO} \log_\lambda(\mathfrak{F}_{s_2}(z))\dif{P(z)} \\
& = \int_{z\in\SO} (\log_\lambda(\mathfrak{F}_{s_1}(z))-\log_\lambda(\mathfrak{F}_{s_2}(z)))\dif{P(z)}\,.
\end{align*}
Now, since $\lambda$ is a fixed point of the deformation in $s$ and since the deformation is continuous in $s$, $\mathfrak{F}_{s_1}(z)$ and $\mathfrak{F}_{s_2}(z)$ (for sufficiently small $|s_1-s_2|$) are sufficiently close on the circle and on the same "side" of the cut so that the algebraic rule of the logarithm holds without the mod $2\pi \ii$. Hence \begin{align*}
\log_\lambda(\mathfrak{F}_{s_1}(U(1))) - \log_\lambda(\mathfrak{F}_{s_2}(U(1))) & = \int_{z\in\SO} \log_\lambda(\mathfrak{F}_{s_1}(z)(\mathfrak{F}_{s_2}(z))^{-1})\dif{P(z)} \\ & = \log_\lambda(\mathfrak{F}_{s_1}(U(1))(\mathfrak{F}_{s_2}(U(1)))^{-1})\,.
\end{align*} This gives
\begin{align*}
	\mathfrak{F}_{s_1}(U))_\lambda(t)-\mathfrak{F}_{s_2}(U))_\lambda(t) &\equiv \exp(t \log_\lambda(\mathfrak{F}_{s_1}(U(1))))-\exp(t \log_\lambda(\mathfrak{F}_{s_2}(U(1)))) \\ 
	& = \mathfrak{F}_{s_1}(U))^\lambda(t) (\Id - \ee^{t \log_\lambda(\Id+(\mathfrak{F}_{s_2}(U(1)) (\mathfrak{F}_{s_1}(U(1)))^{-1}-\Id)})\,.
\end{align*}

We thus find that 
\begin{align*}
	\slim_{\ve\to0}\frac{1}{\ve}(\mathfrak{F}_{s+\ve}(U))_\lambda(t)-\mathfrak{F}_{s}(U))_\lambda(t)) = t  \mathfrak{F}_{s}(U))_\lambda(t)  (\partial_s \mathfrak{F}_{s})(U(1)) (\mathfrak{F}_{s}(U(1)))^{-1} \,,
\end{align*}
which is weakly-local, as $\lambda$ always falls within the mobility gap of $\mathfrak{F}_{s}(U(1)))$. For the time derivative we get similar formulas and following the same argument as above, we find that $|W(\mathfrak{F}_{s_1}(U))_\lambda)-W(\mathfrak{F}_{s_2}(U)_\lambda)|$ can also be made arbitrarily small.
\end{proof}

\begin{lem}\label{lem:W continuity estimate}
	For any two unitary maps $A,B:[0,1]\to\mathcal{U}(\mathcal{H})$ which are differentiable, whose derivatives are bounded too, and which are weakly-local, we have \begin{align}|W(A)-W(B)| &\leq \sup_{[0,1]} \|T_1(A-B)\|_1 + \|T_2(\dot{A}-\dot{B})\|_1 +\nonumber \\ 
	& \quad +\sup_{\alpha,\beta} (\|T_{3\alpha}(A-B)_{,\beta}\|_1 + \|(A-B)^\ast_{,\alpha}T_{4\beta}\|_1+\label{eq:continuity of Cartan-Mauerer}\\
	& \qquad \qquad +\|(A-B)^\ast_{,\alpha}(A-B)_{,\beta}\|_1) \nonumber \end{align} where $T_1,T_2$ are some (time-dependent) trace class operators depending on $A,B$ their derivatives w.r.t. time and their spatial derivatives, the supremum over $\alpha,\beta$ is over the two possibilities where $\alpha\neq\beta$. Then $T_{3\alpha},T_{4\beta}$ is a weakly-local operator confined in the $\alpha,\beta$ direction.
	\begin{proof}
		We start from \cref{eq:Cartan-Mauerer-form} which says $$W(A) = -\frac{1}{2}\int_0^1 \tr  \ve_{\alpha\beta}\dot{A} A^\ast A_{,\alpha}A^\ast A_{,\beta}A^\ast$$ to get \begin{align*}
		|W(A)-W(B)| &\leq \frac{1}{2}\sup_{[0,1]} \|  \ve_{\alpha\beta}(\dot{A} A^\ast A_{,\alpha}A^\ast A_{,\beta}A^\ast-\dot{B} B^\ast B_{,\alpha}B^\ast B_{,\beta}B^\ast)\|_1 \\
		&\leq \frac{1}{2}\sup_{\alpha,\beta,[0,1]} \|  \dot{A} A^\ast A_{,\alpha}A^\ast A_{,\beta}A^\ast-\dot{B} B^\ast B_{,\alpha}B^\ast B_{,\beta}B^\ast\|_1 \\
		& \leq \frac{1}{2}\sup_{\alpha,\beta,[0,1]}( \|\dot{A} A^\ast (A_{,\alpha}A^\ast A_{,\beta}A^\ast- B_{,\alpha}B^\ast B_{,\beta}B^\ast)\|_1 + \\
		& \qquad + \|(\dot{A} A^\ast-\dot{B} B^\ast)B_{,\alpha}B^\ast B_{,\beta}B^\ast \|_1) \\ 
		& \leq \frac{1}{2}\sup_{\alpha,\beta,[0,1]}( \|\dot{A}\|\| A_{,\alpha}A^\ast A_{,\beta}A^\ast- B_{,\alpha}B^\ast B_{,\beta}B^\ast\|_1+\\ 
		& \qquad + \|\dot{A}\|\|(A-B)B_{,\alpha}B^\ast B_{,\beta}B^\ast\|_1+\|(\dot{A}-\dot{B})B^\ast B_{,\alpha}B^\ast B_{,\beta} \|_1) \,.
		\end{align*}
		The supremum is over all times in $[0,1]$ and all $\alpha,\beta$ equal to $1,2$ (without $\alpha=\beta$). 
		
		We concentrate on the term $\| A_{,\alpha}A^\ast A_{,\beta}A^\ast- B_{,\alpha}B^\ast B_{,\beta}B^\ast\|_1$ since the two other terms are in their final desired form. Because $A,B$ are unitary we have $A_{,\alpha}A^\ast = -A A^\ast_{,\alpha}$ so that \begin{align*}
		A_{,\alpha}A^\ast A_{,\beta}A^\ast- B_{,\alpha}B^\ast B_{,\beta}B^\ast &= -A A^\ast_{,\alpha} A_{,\beta}A^\ast+ B B^\ast_{,\alpha} B_{,\beta}B^\ast \\
		& = (B-A) A^\ast_{,\alpha} A_{,\beta}A^\ast - \\
		& -B (A^\ast_{,\alpha} A_{,\beta}-B^\ast_{,\alpha} B_{,\beta})A^\ast+ \\
		& +B B^\ast_{,\alpha} B_{,\beta}(B-A)^\ast \,.
		\end{align*}
		Only the middle line is not in the form we want, so that we write, \begin{align*}A^\ast_{,\alpha} A_{,\beta}-B^\ast_{,\alpha} B_{,\beta} = A^\ast_{,\alpha}(A-B)_{,\beta} + (A-B)^\ast_{,\alpha}A_{\beta}-(A-B)^\ast_{,\alpha}(A-B)_{,\beta}\,. \end{align*}		
		
	\end{proof}
\end{lem}

\appendix

\section{Appendix}

\subsection{Floquet's RAGE \label{app:RAGE}}
In this section we prove that our deterministic dynamical localization assumption implies pure point spectrum (so that it's not necessary to also have the latter as an assumption). This entails importing the analysis of the RAGE theorem to the unitary Floquet case. Most of this was already done in \cite{HamzaJoyeStolz09} but since there it is written for a probabilistic model and we insist in this paper rather on deterministic assumptions and statements (compare our deterministic \cref{eq:U is localized} with their probabilistic \cite{HamzaJoyeStolz09}, eq-n (3.1)), and also in order to setup the notation for our important \cref{lem:The stretch-construction has pure point spectrum}, we included the proof here as succinctly as possible. 

Within this section, let a unitary $U\in\mathcal{B}(\ell^{2}(\mathbb{Z}^{d})\otimes\mathbb{C}^{N})$
be given such that it is \emph{localized}. For our purposes it is
enough to make the following 
\begin{defn}
	$U$ is deterministically dynamically localized\emph{ in the interval
		$I\subseteq S^{1}$} iff there is some $\mu>0$ such that for any
	$\ve>0$ there is a $0<C_{\ve}<\infty$ such that
	the following holds
	\begin{align}
	\sup_{n\in\mathbb{N}}\sum_{x,\,y\in\mathbb{Z}^{d}}\norm{\left\langle \delta_{x},\,U^{n}\chi_{I}(U)\delta_{y}\right\rangle }e^{\mu\norm{x-y}-\ve\norm{x}} & =C_{\ve}\label{eq:U is localized}
	\end{align}
\end{defn}

\begin{lem}
	\label{lem:Discrete-Wiener}(Discrete Wiener) Let $\mu$ be a complex
	measure on $S^{1}$. For $m\in\mathbb{N}$, we define its $m$th complex
	moment as $\mu_{m} := \int_{z\in S^{1}}z^{m}\dif{\mu(z)}$. Then $
		\lim_{n\to\infty}\frac{1}{n}\sum_{m=1}^{n}\left|\mu_{m}\right|^{2}  =  \sum_{z\in S^{1}}\left|\mu(\Set{z})\right|^{2}
$ that is, the RHS gives the pure point part of $\left|\mu(S^{1})\right|^{2}$.
	\begin{proof}
		We have 
		\begin{align*}
			\frac{1}{n}\sum_{m=1}^{n}\left|\mu_{m}\right|^{2} & \equiv  \frac{1}{n}\sum_{m=1}^{n}\int_{z\in S^{1}}z^{m}\dif{\mu(z)}\int_{w\in S^{1}}\overline{w}^{m}\dif{\overline{\mu}(w)}
			 =  \frac{1}{n}\sum_{m=1}^{n}\int_{w\in S^{1}}\int_{z\in S^{1}}(z\overline{w})^{m}\dif{\mu(z)}\dif{\overline{\mu}(w)}\\
			& =  \int_{w\in S^{1}}\int_{z\in S^{1}}\frac{1}{n}\sum_{m=1}^{n}(z\overline{w})^{m}\dif{\mu(z)}\dif{\overline{\mu}(w)}
		\end{align*}
		Note that the sequence of functions $\Set{S^{1}\ni z\mapsto\frac{1}{n}\sum_{m=1}^{n}z^{m}}_{n\in\mathbb{N}}$
		is uniformly bounded by $1$ and converges pointwise to $\delta(\cdot-1)$.
		We may thus use the dominated convergence theorem to find 
		\begin{align*}
			\lim_{n\to\infty}\frac{1}{n}\sum_{m=1}^{n}\left|\mu_{m}\right|^{2} & =  \int_{w\in S^{1}}\int_{z\in S^{1}}\delta(z\overline{w}-1)\dif{\mu(z)}\dif{\overline{\mu}(w)}
			 =  \int_{z\in S^{1}}\dif{\mu(z)}\overline{\mu(\Set{z})}
			 =  \sum_{z\in S^{1}}\left|\mu(\Set{z})\right|^{2}\,.
		\end{align*}
		
	\end{proof}
\end{lem}

\begin{lem}
	\label{lem:Consequence of Wiener for the continuous spectrum}Let
	$U$ be unitary and $K$ compact. Then 
	\begin{align*}
		\lim_{n\to\infty}\frac{1}{n}\sum_{m=1}^{n}\norm{KU^{m}\psi}^{2} & =  0
	\end{align*}
	for all $\psi\in\mathcal{H}^{c}$, the continuous part of the Hilbert
	space for $U$.
	\begin{proof}
		This is \cite[Lemma 2.7]{AizenmanWarzel2016} in our setting of discrete rather than continuous time. We thus omit the proof. 
%
	\end{proof}
\end{lem}

\begin{thm}
	\label{thm:Floquet's RAGE}(Unitary RAGE) Let $U$ be unitary and
	$\Set{A_{L}}_{L}$ be a sequence of compact operators strongly converging
	to $\mathds{1}$. Then 
	\begin{align*}
		\mathcal{H}^{c} & =  \Set{\psi\in\mathcal{H}|\lim_{L\to\infty}\lim_{n\to\infty}\frac{1}{n}\sum_{m=1}^{n}\norm{A_{L}U^{n}\psi}^{2}=0}\,,
	\end{align*}
	and 
	\begin{align*}
		\mathcal{H}^{p} & =  \Set{\psi\in\mathcal{H}|\lim_{L\to\infty}\sup_{n\in\mathbb{N}}\norm{(\mathds{1}-A_{L})U^{n}\psi}=0}\,.
	\end{align*}
	\begin{proof}
		This is \cite[Theorem 2.6]{AizenmanWarzel2016} in our setting of discrete rather than continuous time, but the same proof goes through with very slight modifications.
	\end{proof}
\end{thm}

The following theorem and the remark after it are the reason for this section.

\begin{thm}
	\label{thm:dyn loc implies spec loc}(Deterministic dynamical localization
	implies spectral localization) If $U$ is deterministically dynamically
	localized in the interval $I$ then it has pure point spectrum within
	that interval, that is, 
	\begin{align*}
		\sigma(U)\cap I & =  \sigma_{pp}(U)\cap I
	\end{align*}
	\begin{proof}
		Since $\Set{\delta_{x}}_{x\in\mathbb{Z}^{d}}$ is an ONB for $\mathcal{H}$,
		and we want to show that $\chi_{I}(U)\mathcal{H}\subseteq\mathcal{H}^{p}$,
		let $y\in\mathbb{Z}^{d}$ be given. We claim that $\chi_{I}(U)\delta_{y}\in\mathcal{H}^{p}$.
		Let $A_{L}$ be the projection onto a box of total volume $(2L+1)^{d}$ centered about the origin
		of $\mathbb{Z}^{d}$. Using
		\cref{thm:Floquet's RAGE} it suffices to show 
		\begin{align*}
			\lim_{L\to\infty}\sup_{n\in\mathbb{N}}\norm{A_{L}^{\perp}U^{n}\chi_{I}(U)\delta_{y}} & =  0\,.
		\end{align*}
		By \cref{eq:U is localized} we have \emph{for any $n\in\mathbb{N}$,}
		\begin{align*}
			\sum_{x,\,y\in\mathbb{Z}^{d}}\norm{\left\langle \delta_{x},\,U^{n}\chi_{I}(U)\delta_{y}\right\rangle }e^{\mu\norm{x-y}-\ve\norm{x}} & \leq  C_{\ve}\,.
		\end{align*}
		This in turn implies that 
		\begin{align*}
			\norm{\left\langle \delta_{x},\,U^{n}\chi_{I}(U)\delta_{y}\right\rangle }e^{\mu\norm{x-y}-\ve\norm{x}} & \leq  \sum_{x',y'}\norm{\left\langle \delta_{x'},\,U^{n}\chi_{I}(U)\delta_{y'}\right\rangle }e^{\mu\norm{x'-y'}-\ve\norm{x'}}\\
			& \leq  C_{\ve}
		\end{align*}
		since all terms are positive. Hence, $\norm{\left\langle \delta_{x},\,U^{n}\chi_{I}(U)\delta_{y}\right\rangle }  \leq  C_{\ve}e^{-\mu\norm{x-y}+\ve\norm{x}}$
		uniformly in $n$.
		
		Now we have
		\begin{align*}
			\norm{A_{L}^{\perp}U^{n}\chi_{I}(U)\delta_{y}}^{2} &  =  \sum_{x\in\mathbb{Z}^{d}:\norm{x}>L}\norm{\left\langle \delta_{x},\,U^{n}\chi_{I}(U)\delta_{y}\right\rangle }^{2}\\
			&   (\text{Using }\norm{\left\langle \delta_{x},\,U^{n}\chi_{I}(U)\delta_{y}\right\rangle }\leq1)\\
			& \leq  \sum_{x\in\mathbb{Z}^{d}:\norm{x}>L}\norm{\left\langle \delta_{x},\,U^{n}\chi_{I}(U)\delta_{y}\right\rangle } \leq  \sum_{x\in\mathbb{Z}^{d}:\norm{x}>L}C_{\ve}e^{-\mu\norm{x-y}+\ve\norm{x}}\,.
		\end{align*}
		Hence since the square root is monotone increasing and continuous, and using $\sqrt{a+b}\leq\sqrt{a}+\sqrt{b}$, we find
		\begin{align*}
			\norm{A_{L}^{\perp}U^{n}\chi_{I}(U)\delta_{y}} & \leq \sqrt{C_{\ve}}\sum_{x\in\mathbb{Z}^{d}:\norm{x}>L}e^{-\frac{1}{2}\mu\norm{x-y}+\frac{1}{2}\ve\norm{x}}
		\end{align*}
		for any $n\in\mathbb{N}$ so that taking the supremum on both sides
		(redundant on the RHS) and then the limit $L\to\infty$ we get zero indeed.
		This follows because (for $\ve<\mu$) $e^{-\frac{1}{2}\mu\norm{x-y}+\frac{1}{2}\ve\norm{x}}$
		is summable in $x$, and hence taking the limit $L\to\infty$ gives
		zero.
	\end{proof}
\end{thm}

\begin{lem}\label{lem:The stretch-construction has pure point spectrum}
	(The stretch-construction and pure point spectrum) Let $U$ be such that $\sigma(U)\cap I$ is pure point and $\sigma(U)\cap I^{c}$
	is some mixture of pure point and continuous spectrum. Define 
	\begin{align*}
		V & =  \chi_{I}(U)f(U)+\chi_{I^{c}}(U)\,.
	\end{align*}
	where $f:S^{1}\to S^{1}$ has a range which is the entire circle. Then $\sigma(V)=\sigma_{pp}(V)$. 
	
\end{lem}

We note that in our application of the stretch-function, strictly-speaking, this lemma could be avoided since $F_\Delta \in B_1(\Delta)$ so that $V(1)$ is actually dynamically-localized as in \cref{def:mobility gap} on $\SO\setminus\{1\}$, and thus one could invoke \cref{thm:dyn loc implies spec loc} to conclude $\sigma(V(1))=\sigma_{pp}(V(1))$. However, the proof below proceeds directly without making an assumption of dynamical localization on $U$, but rather, only on its spectral type within $I$. 

\begin{proof}
	We have by \cref{thm:Floquet's RAGE}, for any $\psi\in\mathcal{H}$
	\begin{align*}
	\norm{(\mathds{1}-A_{L})V^{n}\psi} & =  \norm{(\mathds{1}-A_{L})(\chi_{I}(U)f(U)+\chi_{I^{c}}(U))^{n}\psi}\\
	&   (\text{By projections being orthogonal})\\
	& =  \norm{(\mathds{1}-A_{L})(\chi_{I}(U)f(U)^{n}\psi+\chi_{I^{c}}(U)\psi)}\\
	& \leq  \norm{(\mathds{1}-A_{L})f(U)^{n}\chi_{I}(U)\psi}+\norm{(\mathds{1}-A_{L})\chi_{I^{c}}(U)\psi}
	\end{align*}
	Now in general we may write $\psi=\psi_{1}+\psi_{2}$ with $\psi_{1}\in\chi_{I}(U)$
	and $\psi_{2}\in\chi_{I^{c}}(U)$. Taking the supremum
	and limit of both sides, using the fact that the supremum of a sum
	is smaller than the sum of supremums, we find 
	\begin{align*}
	\lim_{L\to\infty}\sup_{n\in\mathbb{N}}\norm{(\mathds{1}-A_{L})V^{n}\psi} & \leq  \lim_{L\to\infty}\sup_{n\in\mathbb{N}}\norm{(\mathds{1}-A_{L})f(U)^{n}\psi_{1}}+\underbrace{\lim_{L\to\infty}\norm{(\mathds{1}-A_{L})\psi_{2}}}_{=0}
	\end{align*}
	We know $\psi_{1}\in\mathcal{H}_{U}^{p}$ by the assumption on $I$.
	That means that either it is an eigenvector of $U$ with eigenvalue
	$\lambda$ or it is in the closure of the set of eigenvalues of $U$.
	In the former case we have $f(U)\psi_{1}=f(\lambda)\psi_{1}$
	whence $\psi_{1}\in\mathcal{H}_{f(U)}^{p}$ so that $\lim_{L\to\infty}\sup_{n\in\mathbb{N}}\norm{(\mathds{1}-A_{L})f(U)^{n}\psi_{1}}=0$
	by \cref{thm:Floquet's RAGE}. Otherwise, for any $\ve>0$
	there is some $\psi_{\ve}\in\mathcal{H}$ such that $\psi_{\ve}$
	\emph{is} an eigenvector of $U$ (with eigenvalue $\lambda_{\ve}$)
	and $\norm{\psi_{1}-\psi_{\ve}}<\ve$. Then 
	\begin{align*}
	\norm{(\mathds{1}-A_{L})f(U)^{n}\psi_{1}} & \leq  \norm{(\mathds{1}-A_{L})f(U)^{n}\psi_{\ve}}+\norm{(\mathds{1}-A_{L})f(U)^{n}(\psi_{1}-\psi_{\ve})}\,.
	\end{align*}
	When taking the supremum and the limit, the first term will tend to
	zero as was just remarked. Thus let us concentrate on the second term:
	\begin{align*}
	\norm{(\mathds{1}-A_{L})f(U)^{n}(\psi_{1}-\psi_{\ve})} & \leq  \ve(1+\norm{A_{L}})\norm{f(U)^{n}}\\
	& \leq  \ve(1+\norm{A_{L}})\underbrace{\sup_{z}\left|f(z)^{n}\right|}_{\leq1}\\
	& \leq  2\ve\,.
	\end{align*}
	Since $\ve>0$ was arbitrary we find the result.
\end{proof}

\subsection{Helffer-Sj\"ostrand formula for unitary operators \label{app:HJ_formula}}

Helffer-Sj\"ostrand formula extends holomorphic functional calculus to smooth functions. It was developed for Hermitian operators but can be easily adapted to unitaries, with the simplification that the latter are always bounded. A formula was already proposed in \cite{Mbarek15} for functions on $\SO \setminus \{1\}$ and based on Cayley transformation. Here we provide another proof for any smooth function on $\SO$ using a conformal mapping.

\begin{lem}\label{lem:HJ_unitary}
	Let $f: \SO \rightarrow \mathbb C$ be a smooth function. There exists a quasi-analytic extension $\tilde f : \mathbb C \rightarrow \mathbb C$, \textit{i.e.} $\tilde f|_{\SO} = f$ and $\partial_{\bar z} \tilde f|_{\SO} =0$, such that for any unitary operator $U$
	\begin{align}\label{eq:HJ_unitary}
		f(U) &= \dfrac{1}{2\pi \ii}\int_{\mathbb C} (\partial_{\bar z} \tilde f(z)) (z-U)^{-1} \dd z \dd \bar z
	\end{align}
	Moreover $\tilde f$ is compactly supported around $\SO$ and satisfies $|\partial_{\bar z} \tilde f|\leq C | |z|-1 |^N$ for any $N \geq 2$.
	\begin{proof}
		Any function $f: \SO \rightarrow \mathbb C$ can be equivalently described by a periodic function $g: \mathbb R \rightarrow \mathbb C$, through the conformal mapping $w \mapsto z=\ee^{\ii w}$ by $g(w) = f(\ee^{\ii w})$, satisfying $g(w+2\pi) = g(w)$ by construction. This bijective mapping extends to the the annulus $\mathcal A_r$ where $\ee^{-r} < |z| < \ee^r$ corresponding to the strip $-r < \Im(w) <r$. In both cases the smoothness of $f$ and $g$ are the same. Let $\chi : \mathbb R \rightarrow \mathbb C$ be a smooth function supported in $(-r,r)$ and with $\chi(x) = 1$ near 0.  
		On the real line, we know from Ref. \cite{HunzikerSigal00} that for $N \geq 2$
		\begin{align}
		\tilde g(\theta,\tau) &= \sum_{k=0}^{N-1} g^{(k)}(\theta) \dfrac{(\ii \tau)^k}{k!}\chi(\tau)
		\end{align}
	is a quasi-analytic extension of $g$ on the strip, namely $\tilde g(\theta,0) = g(\theta)$ and $\partial_{\bar{w}} g|_{\tau=0} = 0$, for $w = \theta + \ii \tau$ and $\partial_{\bar{w}} = 1/2(\partial_\theta + \ii \partial_\tau)$. Moreover, $|\partial_{\bar{w}}g| \leq C |\tau|^N$. We claim that $\tilde f(z=\ee^{\ii (\theta +\ii \tau)}) := \tilde g(\theta,\tau)$ is a quasi-analytic extension of $f$ on the annulus. Indeed $\tilde f$ coincides with $f$ on $\SO$ and 
	\begin{align}
	\partial_{\bar{w}} \tilde g &= \partial_{\bar{w}}(\overline{\ee^{\ii w}}) \partial_{\bar z} \tilde f = -\ii \bar z \partial_{\bar z} \tilde f 
	\end{align}
	so that $\partial_{\bar z}\tilde f|_{\tau=0}=0$. Moreover on the annulus one has $\ee^{-r} < |-\ii \bar z| <\ee^{r}$ and $|\ln x|\leq \ee^{r} |x-1|$ for $x \in (\ee^{-r},\ee^{r})$ applied to $x=|z|= \ee^{-\tau}$ we infer $|\tau| \leq C ||z|-1|$ so that
	\begin{align}
	|\partial_{\bar z} \tilde f| &\leq C ||z|-1|^N
	\end{align}
	with a different constant $C$. With the fact that $\norm{(z-U)^{-1}} \leq | |z| -1|^{-1}$ for a unitary $U$ we deduce that the integral in \eqref{eq:HJ_unitary} is absolutely convergent in norm. Then we claim that for $z_0 \in \mathbb S^1$
	\begin{align}
	f(z_0) &= \dfrac{1}{2\pi}\int_{\mathbb C} (\partial_{\bar z} \tilde f(z)) (z-z_0)^{-1} \dd z \dd \bar z
	\end{align}
	The integral is reduced to the annulus $\mathcal A_r$ since $\tilde f$ is supported inside it and has to be understood as an improper integral on $\mathcal A_r \setminus \mathcal A_\ve$ when $\epsilon \rightarrow 0$. The equality follows by \cite[Cor. 2.3]{Mbarek15}, and \eqref{eq:HJ_unitary} follows by the functional calculus.
	\end{proof}
\end{lem}

\begin{cor}\label{cor:smooth_f_is_weakly_local} The smooth functional calculus of an exponentially local unitary is polynomially local.
	
	\begin{proof}
		This is a direct consequence of the Helffer-Sj\"ostrand formula \eqref{eq:HJ_unitary}, the fact that $\tilde f$ is smooth and compactly supported, and Combes-Thomas estimate \cite{Combes_Thomas_1973} (\cite{HamzaJoyeStolz09} in the context of unitaries): if $U$ is local then it exists $0<C<\infty$ such that
		\begin{align}
		|R_U(z)|_{x,y} \leq \dfrac{C}{||z|-1|} \ee^{-\mu(z)\norm{x-y} }
		\end{align}
		for $\mu>0$ small enough. For example one can take $\mu(z) = c ||z| -1|$ as in \cite{Elbau_Graf_2002}. According to \cref{lem:HJ_unitary} the quasi-analytic extension of $f$ satisfies $|\partial_{\bar z} \tilde f(z) | \leq C ||z|-1|^N$ for $N \geq 2$ so that
		\begin{align}
		|f(U)|_{x,y} \leq \dfrac{1}{2\pi} \int \dd z \dd \bar z |\partial_{\bar z} \tilde f(z) | |R_U(z)|_{x,y} \leq C_N (1+c\norm{x-y})^{-N}
		\end{align}
	\end{proof}
\end{cor}
\subsection{Convergence properties of weakly-local operators\label{app:convergence_lemma}}
\begin{lem}
	If $A_n \to A$ strongly within the star-algebra of weakly-local operators then $\partial_j A_n \to \partial_j A$ strongly within the ideal of weakly-local-and-confined in direction $j$ operators.
	\begin{proof}
		We already know that $\partial_j A_n$ (for all $n$) and $\partial_j A$ are weakly-local-and-confined in direction $j$ by the results of \cref{subsec:the-weakly-local-and-confined-two-sided-ideal}. Now let $\psi\in\mathcal{H}$ be given. We have \begin{align*}
			\|\partial_j A_n \psi-\partial_j A \psi\| &\leq \|\Lambda_{j}(A_{n}-A)\psi-(A_{n}-A)\Lambda_{j}\psi\| \\
			&\leq \|(A_n-A)\psi\|+\|(A_n-A)\Lambda_j\psi\|\\
			&\to0\,.
		\end{align*}
	\end{proof}
\end{lem}
\begin{lem}\label{lem:A_n to A strongly and all weakly local T_alpha confined in alpha}
	If $A_n \to A$ strongly within the ideal of weakly-local-and-confined in direction $1$ operators, all having a uniform both $\nu$ and sufficiently large $\mu$ as in \cref{defn:weakly-local-and-confined-operator}, and $T$ is weakly-local-and-confined in direction $2$, then $TA_n \to TA$ in trace-class norm.
	\begin{proof}
		We have $T A_n = T (1+|X_1|)^{-\mu}(1+\|X\|)^{\nu}(1+\|X\|)^{-\nu}(1+|X_1|)^{\mu} A_n$. WLOG, we also pick $\mu$ such that $T(1+|X_1|)^{-\mu}(1+\|X\|)^{\nu}$ is trace-class, and note that $(1+\|X\|)^{-\nu}(1+|X_1|)^{\mu} A_n\to(1+\|X\|)^{-\nu}(1+|X_1|)^{\mu} A$ strongly. We verify these two statements: \begin{align*}
			\|T(1+|X_1|)^{-\mu}(1+\|X\|)^{\nu}\|_1 &\leq \sum_{xy} \|T_{xy}\| (1+|y_1|)^{-\mu}(1+\|y\|)^{\nu} \\
			& \leq \sum_{xy} C^T_\mu (1+\|x-y\|)^{-\mu}(1+|y_2|)^{-\mu}\\ & \hspace{4cm}(1+\|y\|)^{\nu} (1+|y_1|)^{-\mu}(1+\|y\|)^{\nu} \\
			& < \infty\,.
		\end{align*}For the second statement, let $C_n := A_n - A$ . Then \begin{align*}
&			\|(1+\|X\|)^{-\nu}(1+|X_1|)^{+\mu} C_n\psi\|^2 \\ &\equiv \langle (1+\|X\|)^{-\nu}(1+|X_1|)^{+\mu} C_n\psi, (1+\|X\|)^{-\nu}(1+|X_1|)^{+\mu} C_n\psi\rangle \\
			& = \langle (1+\|X\|)^{-2\nu}(1+|X_1|)^{+2\mu} C_n\psi, C_n\psi\rangle \\
			& \leq \|(1+\|X\|)^{-2\nu}(1+|X_1|)^{+2\mu} C_n\psi\| \|C_n\psi\| \\
			& \leq \|(1+\|X\|)^{-2\nu}(1+|X_1|)^{+2\mu} C_n\| \|C_n\psi\|\,.
		\end{align*}
		The first norm is finite (for each $n$) by \cref{lem:operator-norm-characterization-of-weakly-local-confined-operators} and the second goes to zero because $C_n\to0$ strongly.

		Then we use the result that if $S$ is trace-class and $B_n \to B$ strongly then $S B_n \to S B$ in trace-class norm with $S := T (1+|X_1|)^{-\mu}(1+\|X\|)^{\nu}$ and $B_n := (1+\|X\|)^{-\nu}(1+|X_1|)^{\mu} A_n$.
	\end{proof}
\end{lem}

\begin{lem}\label{lem:A_n to A and B_n to B strongly weakly-local and confined implies trace-class convergence}
	If $A_n \to A,B_n\to B$ strongly within the ideals of weakly-local-and-confined in direction $1$ and $2$ respectively, all having a uniform both $\nu$ and sufficiently large $\mu$ as in \cref{defn:weakly-local-and-confined-operator}, then $A_n B_n \to AB$ in trace-class norm.
	\begin{proof}
		We again write the factorization \begin{align*}
			A_n B_n &= A_n (1+|X_1|)^\mu (1+\|X\|)^{-\nu} \cdot \\ 
					& \cdot (1+|X_1|)^{-\mu} (1+\|X\|)^{2\nu} (1+|X_2|)^{-\mu} \cdot \\
					& \cdot (1+\|X\|)^{-\nu}(1+|X_2|)^\mu B_n \\
					&= A_n (1+|X_1|)^\mu (1+\|x\|)^{-\nu} \cdot (1+|X_1|)^{-\mu/2} (1+\|X\|)^{\nu} (1+|X_2|)^{-\mu/2}\cdot \\ 
					& \cdot (1+|X_1|)^{-\mu/2} (1+\|X\|)^{\nu} (1+|X_2|)^{-\mu/2} \cdot (1+\|X\|)^{-\nu}(1+|X_2|)^\mu B_n\,.
		\end{align*} Now if $\mu$ is chosen sufficiently large, then the last expression is the product of four factors. The first one converges strongly as shown in the lemma above. The second and third are trace class and the fourth also converges strongly. Thus we conclude the statement based on the properties of products of limits and the previous lemma.
	\end{proof}
\end{lem}

\begingroup
\let\itshape\upshape
\printbibliography
\endgroup
\end{document}